\documentclass[preprint, superscriptaddress, amsmath, amssymb, aps, prl, twocolumn,
]{revtex4-2}

\usepackage{graphicx}% Include figure files
\usepackage{dcolumn}% Align table columns on decimal point
\usepackage{bm}% bold math
\usepackage{siunitx}
\usepackage{color}
\usepackage{multirow}
\usepackage{array}
\usepackage{tabularx}
\usepackage{amsmath}
\usepackage{cleveref}
\usepackage[english]{babel}
\usepackage{booktabs}

\definecolor{alex}{rgb}{0.04296875,0.38671875,0.30859375}

\newcommand{\mh}{H\textsubscript{2}}

\begin{document}

\title{Role of anisotropic electronic friction in laser-driven hydrogen recombination on copper}

\author{Alexander Spears}
\affiliation{Department of Chemistry, University of Warwick, Gibbet Hill Road, CV4 7AL Coventry, UK}
\affiliation{University of Vienna, Faculty of Physics, Kolingasse 14-16, Vienna, Austria}
\author{Wojciech G. Stark}
\affiliation{Department of Chemistry, University of Warwick, Gibbet Hill Road, CV4 7AL Coventry, UK}
\affiliation{Department of Chemistry, Imperial College London, White City Campus, Molecular Sciences Research Hub, 82 Wood Ln, London W12 0BZ, UK}

\author{Reinhard J. Maurer}
\email{reinhard.maurer@univie.ac.at}
\affiliation{Department of Chemistry, University of Warwick, Gibbet Hill Road, CV4 7AL Coventry, UK}
\affiliation{University of Vienna, Faculty of Physics, Kolingasse 14-16, Vienna, Austria}
\affiliation{Department of Physics, University of Warwick, Gibbet Hill Road, CV4 7AL Coventry, UK}

\date{\today}

\begin{abstract}
Ultrafast light-driven chemical dynamics at surfaces are governed by energy transfer from excited electrons to vibrational degrees of freedom.
When this nonadiabatic energy transfer is anisotropic, it can lead to dynamical steering effects that affect reaction probabilities or non-thermal final energy distributions in molecules. Here, enabled by a machine-learning-based simulation framework, we compare isotropic and anisotropic models of electronic friction during laser-driven hydrogen evolution on the (111) facet of copper. While anisotropic electronic friction determines the rate of energy transfer into the adsorbate and the fluence dependence of desorption probabilities, final translational, vibrational and rotational energy distributions are mainly governed by the potential energy landscape at the barrier and insensitive to nonadiabatic effects.
\end{abstract}

\maketitle
Adsorption and desorption of molecular hydrogen from metal surfaces are important steps in heterogeneous catalysis and electrochemical processes~\cite{langmuir1912chemi}.
Although these steps are thoroughly studied in thermally activated catalysis, many open questions remain for photocatalytic processes on metal interfaces~\cite{lee2023elect}.
Plasmonic metals such as copper, aluminium or gold are often of particular interest due to their efficiency in converting incident light into electron-hole pair excitations beyond thermal excitation thresholds~\cite{petek2023plasm, campillo2000first}.
These non-thermal excitations rapidly thermalise into a "hot" electron distribution through electron-electron scattering with an effective temperature above the surrounding lattice~\cite{frischkorn2006femto}.
While hot electron thermalisation typically occurs within several fs, electron-phonon scattering occurs over multiple ps~\cite{rethfeld2004times}. Until the energy from electron-hole-pair (EHP) excitations is completely equilibrated with the surrounding lattice, energy can be transferred to adsorbates~\cite{frischkorn2006femto}, driving chemical transformations~\cite{frischkorn2008ultra}.

In ultrafast pump-probe experiments, this non-thermal EHP-mediated energy transfer is signified by transient changes to adsorbate vibrational modes, e.g. the red-shift of the CO stretch vibration frequency on a Cu(100) substrate~\cite{inoue2016disen}.
The anisotropy of coupling between hot electrons and adsorbate vibrations has been shown to affect dynamics in measured and calculated vibrational linewidths~\cite{hertl2026mode, morin1992vibra, gadzuk1984vibra} of adsorbed hydrogen, atomic hydrogen scattering~\cite{box2024room}, and dissociative chemisorption of \mh~\cite{maurer2017mode, stark2025nonad}.
In laser-induced dynamics, isotope effects can be enhanced~\cite{lindner2023femto}, trapping in precursor states can be promoted~\cite{mladineo2025photo}, or direction-specific diffusion can be induced~\cite{gonzalez2025femto}.
In addition, laser-desorbed \mh{} molecules generally have a non-thermal partitioning of energy across vibrational, rotational and translational degrees of freedom~\cite{frischkorn2008ultra}.
However, it is still not clear if this non-equipartition of energy in the molecule results from nonadiabatic effects or from details of the barrier landscape during recombinative desorption.

\begin{figure}
	\centering
	\includegraphics[width=\linewidth]{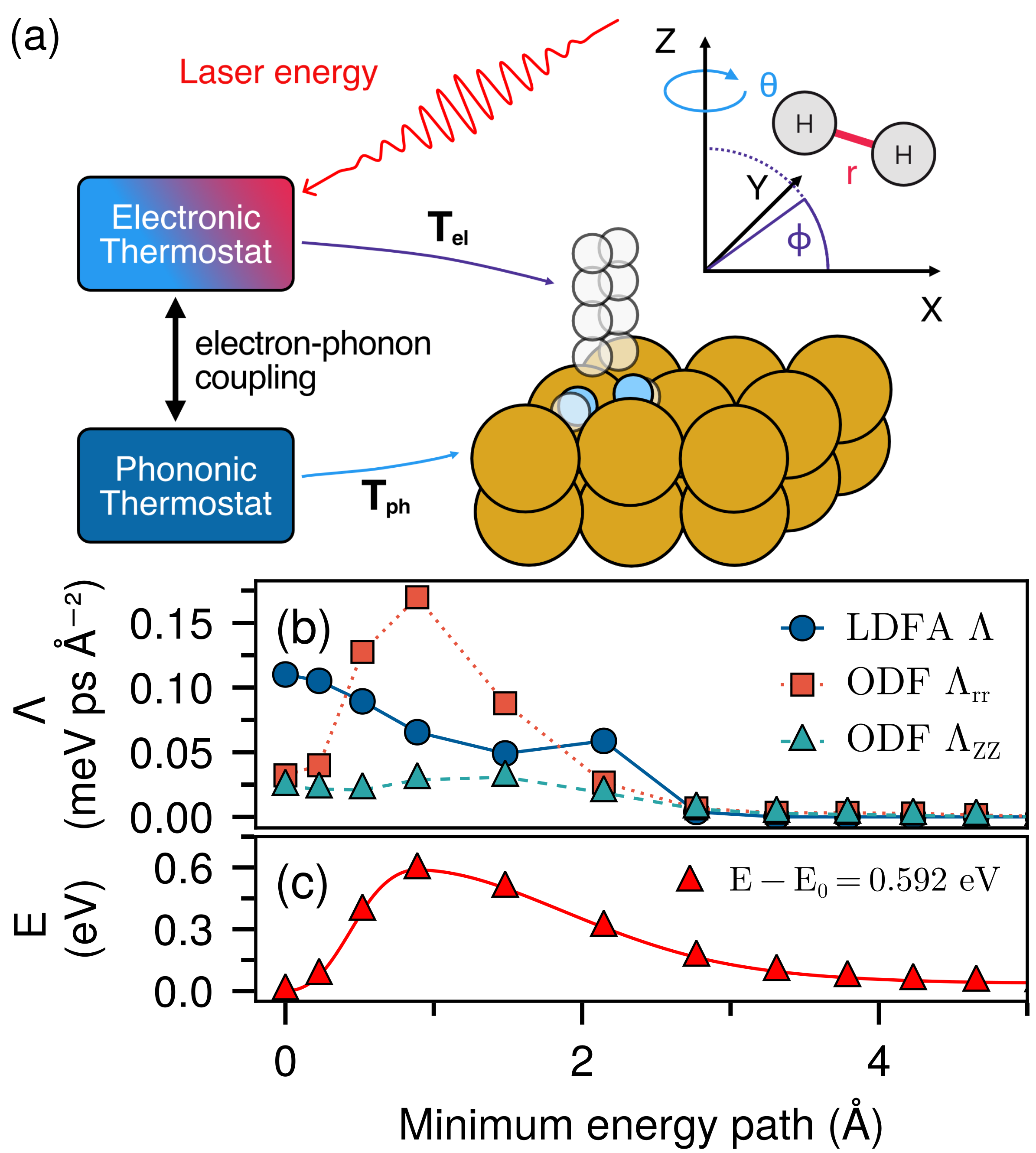}
	\caption{\textbf{(a)} Geometries along the minimum energy path (MEP) for recombinative desorption of H\textsubscript{2} from a Cu(111) surface and a schematic of the electron and phonon bath in the two-temperature model. The transition state geometry is marked in blue. Only the first two of six total layers of the surface slab are shown. \textbf{(b)} Values of electronic friction tensor components along the MEP. \textbf{(c)} Potential energy along the MEP.
    }
	\label{fig-mep-ttm-and-eft-components}
\end{figure}

Classical molecular dynamics simulations impose the Born-Oppenheimer approximation and are unsuitable to describe laser-driven dynamics, as they neglect nonadiabatic electron-nuclear coupling effects~\cite{gonzalez2021quant}.
While a fully \textit{ab-initio} description of electron dynamics triggered by a laser pulse is possible~\cite{schafer2022short}, simulations of the ensuing coupled-electron nuclear dynamics of realistic interfaces are still computationally challenging~\cite{shi2025full}.
A commonly employed mixed quantum-classical approach to study light-driven dynamics at surfaces is the Molecular Dynamics with Electronic Friction (MDEF) method~\cite{head-gordon1995molec, fuchsel2011dissi, luntz2006femto, alducin2019elect}.
In addition to the conservative dynamics governed by the potential energy surface (PES), $V$, the MDEF equation of motion includes nonadiabatic effects as a dissipative force mediated by an electronic friction tensor (EFT), $\Lambda(\mathbf{R})$, and a coupled random force term, $\mathbf{\mathcal{\dot W}}[T(t), t, \Lambda]$:
\begin{equation}\phantomsection\label{eq-mdef}{
    m\mathbf{\ddot R}=-\nabla V-\Lambda(\mathbf{R})\mathbf{\dot R}+\mathbf{\mathcal{\dot W}}(T(t), \Lambda).
}\end{equation}
Light-driven surface dynamics can be simulated by coupling MDEF with the Two-Temperature Model (TTM)~\cite{anisimov1974emiss}, a parametrised  set of coupled heat transfer equations for the electronic ($\mathrm{T_{el}}$) and phononic ($\mathrm{T_{ph}}$) subsystem temperatures~\cite{luntz2005adiac, scholz2016femto} (see \cref{fig-mep-ttm-and-eft-components}a).

Earlier dynamics studies were able to qualitatively match experimental desorption yields and non-thermal energy partitioning for \mh~evolution on Ru(0001) using low-dimensional PESs with static surfaces constructed from Density Functional Theory (DFT)~\cite{fuchsel2011dissi}.
Recently, the emergence of high-dimensional machine learning (ML) potentials has enabled studies that include all degrees of freedom. Simulations of laser-activated CO diffusion on Cu(110)~\cite{gonzalez2025femto} and O diffusion on Pd(111)~\cite{wang2025unvei} have demonstrated how surface and adsorbate excitations cooperatively activate diffusion processes multiple ps after an incident laser pulse. In addition, a study by Lindner \textit{et al} on laser-driven H\textsubscript{2} evolution from Ru(0001)~\cite{lindner2023femto} recently showed how finite-size effects affect desorption yields, calling the accuracy of low-dimensional surrogate models into question.
Virtually all TTM-MDEF simulations in the literature evaluate the EFT with the isotropic Local Density Friction Approximation (LDFA), where friction coefficients are calculated from ion scattering shifts in a homogeneous electron gas with the electron density of the substrate at the position of the adsorbate atom~\cite{juaristi2008role}.

In this Letter, we perform full-dimensional TTM-MDEF simulations of laser-driven \mh{} evolution from a Cu(111) surface at a coverage of 2/9~ML using a tensorial and anisotropic description of orbital-dependent friction (ODF) that is directly calculated from first-principles electron-phonon coupling~\cite{hellsing1984elect, persson1982elect, maurer2016initi, box2023initi}. ODF has been shown to accurately reproduce the mode dependence of vibrational lifetimes~\cite{maurer2016initi, box2024room} and spatially anisotropic energy dissipation effects in reactive scattering of N\textsubscript{2} on Ru(0001)~\cite{shakouri2017accur} and H\textsubscript{2} on Cu~\cite{luntz2009commf}. Previous LDFA-based simulations were able to predict the non-equipartition of energy for \mh{} evolution from Ru(0001) surfaces~\cite{fuchsel2011dissi}. However, LDFA is known to underestimate vibrational lifetimes of H atoms adsorbed on metals~\cite{hertl2026mode,luntz2009commf, maurer2016initi, box2024room}.  Here, we show that the non-equipartition of energy originates dominantly from the shape of the PES, while the rate of recombination and fluence dependence are determined by the employed electronic friction theory.

The TTM-MDEF simulations are enabled by previously reported machine learning models of the PES and the EFT~\cite{stark2025nonad, stark2024bench, stark2023machi} trained on accurate DFT data based on the SRP48 functional~\cite{srp48functional}. As a result, we are able to include all surface degrees of freedom and the effect of laser heating on the phonons. The models were originally developed to describe molecular scattering, but a detailed validation against DFT data for recombinative desorption showed that they are also transferable to laser-assisted desorption (see Supporting Material [SM]~\cite{MyPaper1-SI})\nocite{blum2009initi, pbefunctional, rpbefunctional, sachs2025machi, batatia2022mace, schutt2019unify, schutt2018schne, witt2023acepo, scholz2019vibra, gardner2022nqcdz, software-ase, gerrits2020elecu, larkoski2006numer, wagner2005energ}.

\begin{figure}
	\centering
	\includegraphics[width=\linewidth]{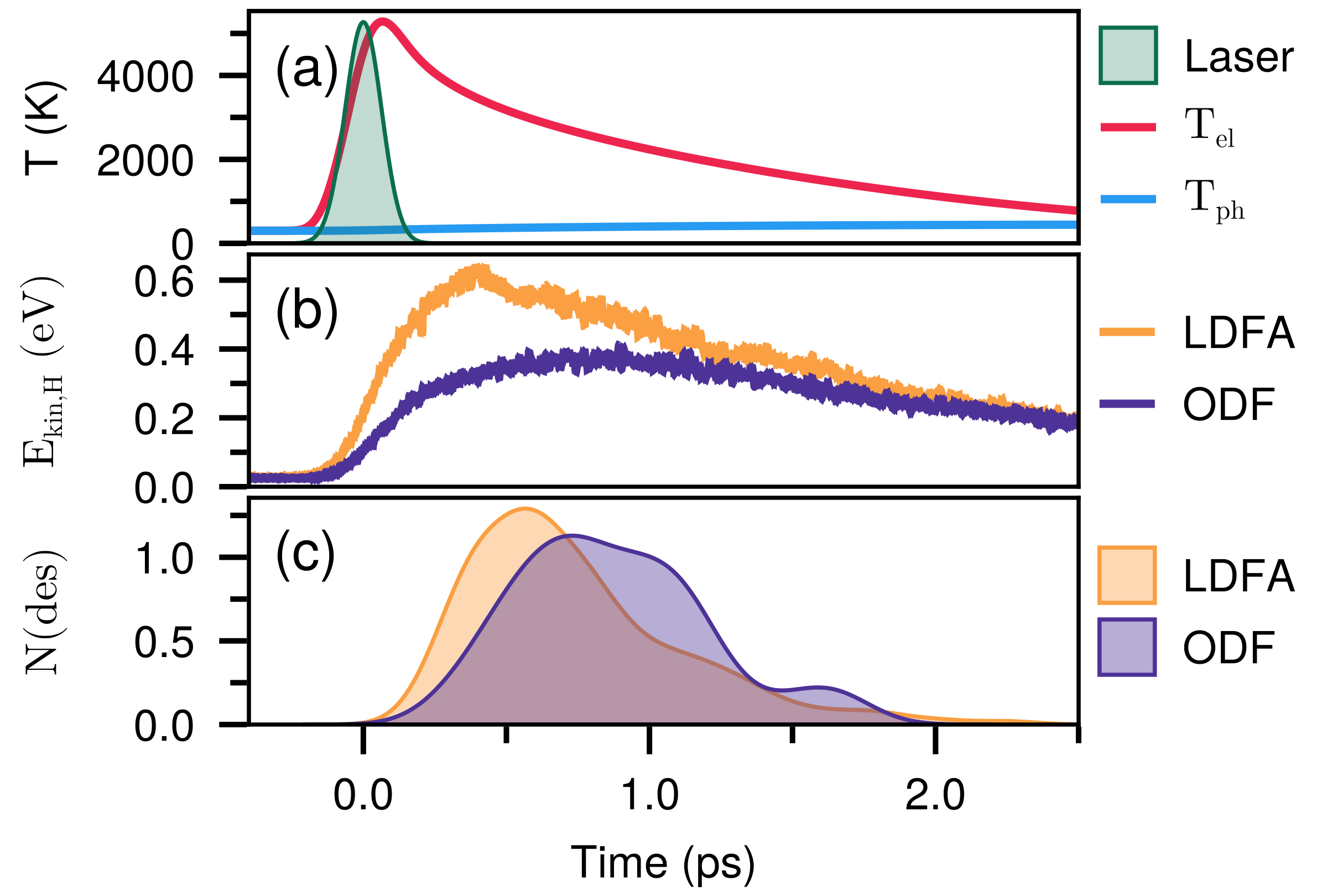}
	\caption{\textbf{(a)}: Temperature profiles for the phonon ($\mathrm{T_{ph}}$) and electron ($\mathrm{T_{el}}$) thermostats in the TTM for a starting temperature of 300~K and a laser fluence of 100~J~m\textsuperscript{-2}. The laser intensity is indicated by the green shaded area and reaches the maximum at $\mathrm{t=0}$. \textbf{(b)}: Average kinetic energies, $E_{\mathrm{kin,H}}$ of adsorbed H atoms over time for both electronic friction approximations. \textbf{(c)}: Normalised distribution of desorption events over time for both electronic friction approximations.}
	\label{fig-ttm-average-energy-and-desorptions}
\end{figure}

\mh{} on Cu(111) exhibits a high desorption barrier [0.592~eV, see \cref{fig-mep-ttm-and-eft-components}c] and strong anisotropy of friction across internal modes of molecular hydrogen during recombination [\cref{fig-mep-ttm-and-eft-components}b].
Upon laser excitation [\cref{fig-ttm-average-energy-and-desorptions}a], the electronic temperature $\mathrm{T_{el}}$ in the metal rapidly rises to over 5000~K within approximately 0.3~ps, while the phononic temperature $\mathrm{T_{ph}}$ rises more slowly.
The rapid increase of $\mathrm{T_{el}}$ leads to activation of hydrogen atom diffusion already within the time span of the laser~\cite{lindner2023femto}.
Adsorbed hydrogen atoms gain kinetic energy twice as quickly on average in LDFA simulations compared to ODF [\cref{fig-ttm-average-energy-and-desorptions}b].
The rate of kinetic energy transfer is determined by the magnitude of the EFT, where, for the chemisorbed hydrogen atoms, the diagonal components of the ODF EFT are roughly half as large as the LDFA friction coefficient (see figures S11 and S12 in the SM).
The ODF is weaker perpendicular to the surface compared to components parallel to the surface (Figure S13 in the SM).
The off-diagonal components of ODF are considerably lower during on-surface diffusion and only reach higher values around the transition to desorption.
The magnitude of ODF only exceeds LDFA in the transition state region and only along the recombination mode (intramolecular stretch), shown as $\Lambda_{rr}$ in \cref{fig-mep-ttm-and-eft-components}b.

The increased kinetic energy of adsorbed hydrogen atoms can eventually lead to recombinative desorption if enough energy has been transferred to cross the recombination barrier.
As a result, the rate of desorption (\cref{fig-ttm-average-energy-and-desorptions}c) reaches a maximum shortly after the maximum kinetic energy transfer into the hydrogen atoms is reached (at 0.39~ps for LDFA, 0.89~ps for ODF).
While the average kinetic energy of H atoms in ODF simulations relaxes more slowly after the laser pulse, we do not find strong  evidence of delaying or dynamical trapping (Figure S15 in the SM).

The choice of electronic friction approximation also leads to almost an order of magnitude difference in desorption yields, with LDFA showing significantly higher reactivity across the range of fluences investigated.
In desorptions induced by multiple electronic transitions~\cite{misewich1992desor}, the desorption probability follows a power law relationship of the form $|A| \mathrm F^n$ with the laser fluence.
This relationship has not been measured experimentally for light-driven H\textsubscript{2} recombination from Cu(111). The TTM-MDEF  simulations find values of $n=3.051\pm 0.201$ for LDFA and $n=2.337\pm 0.197$ for ODF simulations.

\begin{figure}[]
	\centering
	\includegraphics[width = \linewidth]{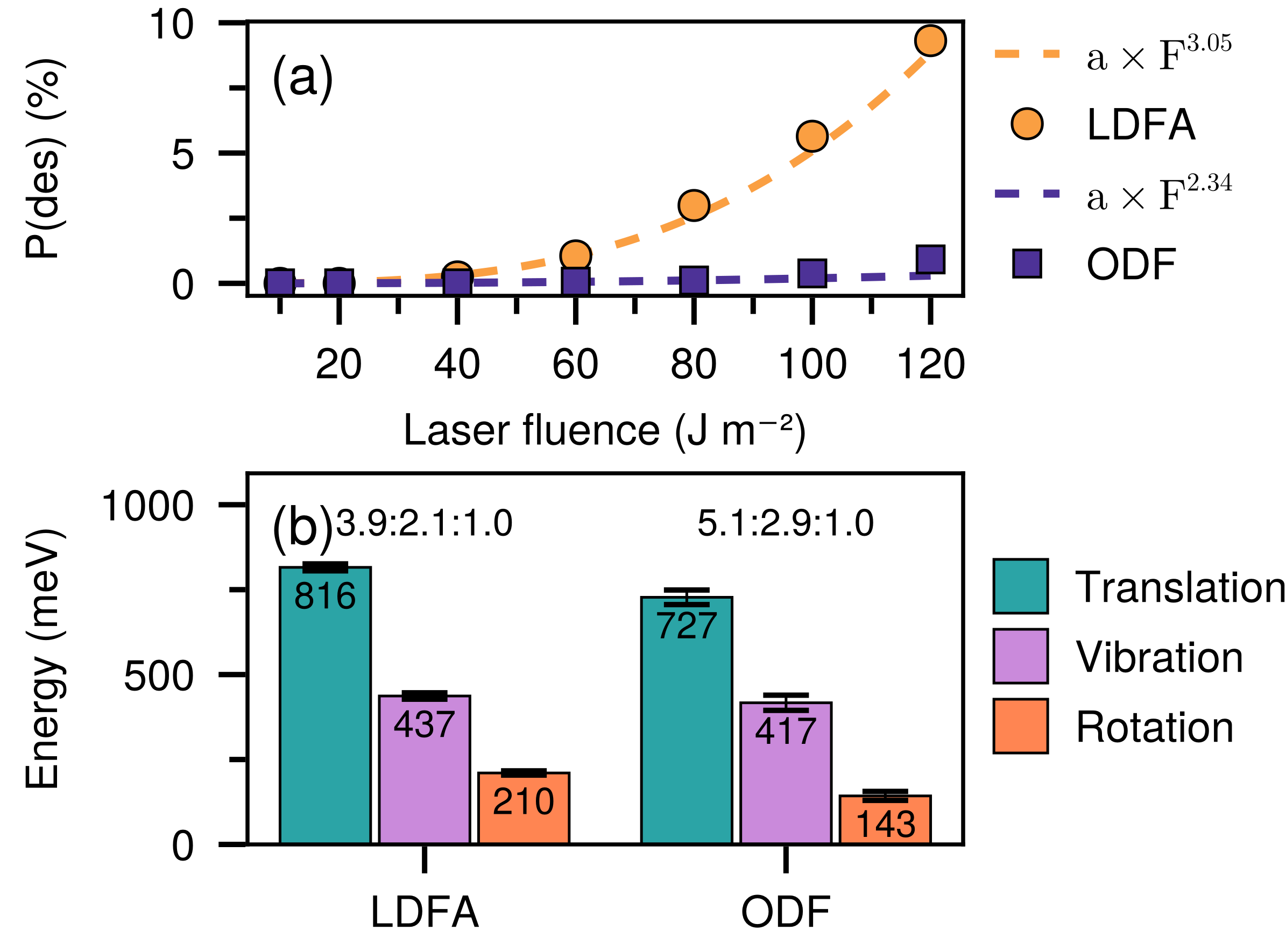}
	\caption{\textbf{(a)}: Single-shot desorption yields for hydrogen on a Cu(111) surface as a function of absorbed laser fluence for simulations with LDFA and ODF. Statistical uncertainties from bootstrapping analysis are indicated where they are larger than the markers used. \textbf{(b)}: Mean internal energies of desorbing H$_2$ molecules determined from LDFA and ODF simulations for a Cu(111) surface with an initial temperature of 100 K and a laser fluence of 120 J m$^{-2}$.}
	\label{fig-probability-and-partitioning}
\end{figure}

\mh{} molecules formed through laser-driven desorption are known to desorb translationally hot, and with a preference for certain rotational and vibrational states, however it is unclear whether the topology of the PES or the energy transfer due to electronic friction (or both) dominate this behaviour~\cite{frischkorn2008ultra, fuchsel2013quant}.
Analysis of the translational, vibrational and rotational energy of desorbing \mh{} molecules in our simulations reveals only subtle differences in the energy partitioning between the two friction models [\cref{fig-probability-and-partitioning}b].
In ODF simulations, \mh{} molecules desorb with lower overall energy than in LDFA simulations.
However, the absolute vibrational energy remains roughly equal for both methods.
This may be a weak indication of the stronger nonadiabatic coupling along the bond axis for ODF [\cref{fig-mep-ttm-and-eft-components}b] \cite{spiering2018testi, stark2025nonad}.
The total vibrational energy is likely somewhat overestimated in both sets of simulations due to the semi-classical Einstein-Brillouin-Keller \cite{larkoski2006numer} analysis rounding to integer quantum states \cite{fuchsel2011dissi}.
Due to the high symmetry and lack of steps or edges on the Cu(111) surface, collision events between hydrogen atoms are most likely to occur when both atoms are at similar heights over the surface, and the MEP favours a concerted motion away from the surface.
These two factors likely discourage cartwheel rotations on desorption from a terrace, leading to desorption with, on average, low rotation energy.
This is slightly more pronounced for ODF, where the component of friction perpendicular to the surface is lower than for LDFA.
Considering the significant difference in magnitude of friction between the two models, the resulting differences in the quantum-state-resolved energy distributions of desorbed molecules are surprisingly small.

As the choice of electronic friction approximation seems to strongly affect desorption probabilities, but only weakly affects the final state distributions of desorbed molecules, we return to the MDEF equation of motion to investigate the mechanism of desorption in more detail. The stochastic equation of motion contains \(\nabla V\), the potential
gradient informing classical dynamics.
The velocity-dependent friction term \(\mathbf \Lambda\left(\partial /\partial t\right)\mathbf{R}\) describes the frictional force caused by nonadiabatic effects, while \(\mathcal{W}(\mathbf \Lambda, \mathrm{T_{el}}, \mathbf{R})\) describes stochastically distributed thermal random noise.

\begin{figure}
	\centering
	\includegraphics[width=\linewidth]{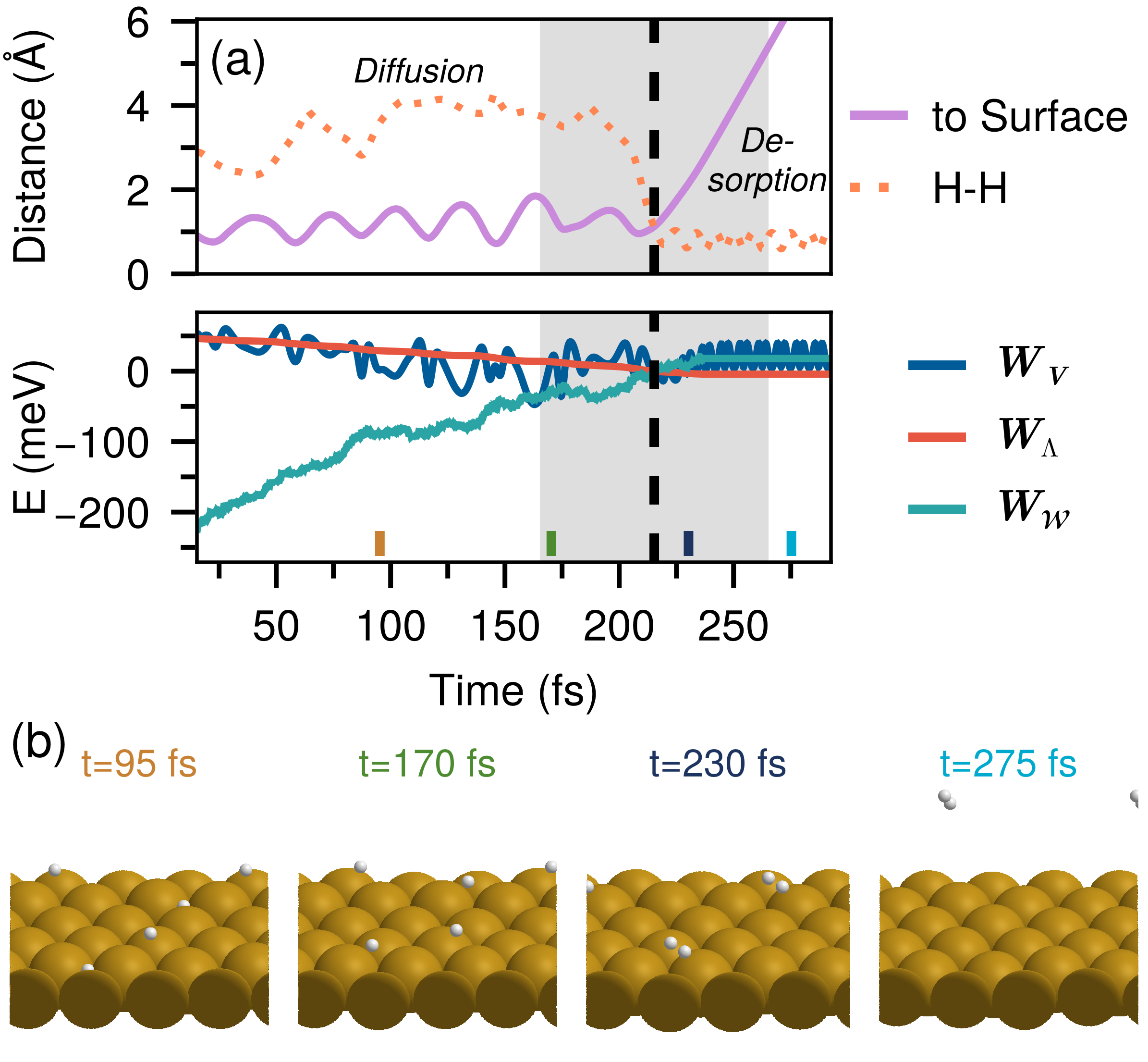}
	\caption{\textbf{(a)}: H-H distance and H\textsubscript{2}-surface distance as a function of time during a trajectory to indicate the two phases of desorption (Diffusion and Desorption, indicated with a grey box). Energy components due to the potential energy $E_V$, friction energy $E_\Lambda$ and thermal fluctuation energy $E_{\mathcal{W}}$ terms for two hydrogen atoms as a function of time during a recombinative desorption event. The EFT is represented with orbital-dependent friction. Equivalent figures for LDFA simulations are in the SM. The energy contributions are shifted to zero at the transition state for clarity. The initial surface temperature was 100~K, the laser fluence was 120~J~m\textsuperscript{-2}. \textbf{b}: Molecular dynamics snapshots throughout the trajectory. }
	\label{fig-energy-accounting-friction}
\end{figure}

For a given simulation, we can determine the work of each of the three force terms described above as a function of time, which allows us to estimate the relevance of each contribution throughout the trajectory, as shown in \cref{fig-energy-accounting-friction}. (Further details on work calculations in the SM~\cite{MyPaper1-SI})
We can distinguish between two stages of a laser-driven desorption event: The "Diffusion" region, where chemisorbed H atoms are mobilised by nonadiabatic energy transfer from EHPs and the "Desorption" region, where two H atoms enter the transition state region of the PES and recombine to H\textsubscript{2}.
In the first region, the thermal fluctuation work $W_{\mathbf \mathcal W}$ continuously transfers energy from hot electrons into the molecule and induces H atom diffusion.
During this stage of the trajectory, $W_{\Lambda}$, the work from the velocity-dependent friction term, dissipates energy out of the system, but at a much lower rate.
In the diffusion regime, where the distance between H atoms is large, off-diagonal elements of the ODF EFT are significantly lower than diagonal components, so differences between LDFA and ODF mostly depend on the magnitudes of diagonal EFT components.
This in turn determines the significant differences in the LDFA and ODF energy transfer rates.

During the desorption stage, i.e. the tens of femtoseconds before and after the onset of desorption [\cref{fig-energy-accounting-friction}b], both contributions are eclipsed by the energy contribution from the PES ($W_V$), consistent with adiabatic, classical dynamics.
Since the time scale between H-H collisions and recombinative desorption lies in the tens of femtoseconds,  there is insufficient time for differences in the magnitude of the LDFA and ODF EFT models to significantly alter the energy distribution of the nascent molecule; in particular, as contributions are effectively zero once the recombination barrier is crossed and the molecule desorbs.
Therefore, the actual process of desorption is insensitive to nonadiabatic effects.
This is consistent with observations on the reverse process: the sticking probability of \mh{} on copper is also insensitive to nonadiabatic effects \cite{spiering2018testi, stark2025nonad}.

In summary, we have compared machine-learning accelerated TTM-MDEF simulations of \mh{} recombination from Cu(111) surfaces using an isotropic and an anisotropic friction model. 
Both models yield differences in energy transfer, diffusion rates and desorption probabilities, yet the desorption process itself is dominated by the barrier and the potential energy landscape. 
The choice of electronic friction approximation primarily determines how quickly the energy barrier to recombinative desorption will be reached, while the translational, vibrational and rotational energy distribution of the desorbing \mh{} molecules is mostly defined by the PES.

LDFA is known to underestimate vibrational lifetimes of adsorbed H atoms \cite{box2024room} and we expect that this also leads to an overestimated desorption yield in our simulations.
However, our comparison only holds in the low coverage regime, where off-diagonal components in ODF are small.
Vibrational lifetimes of adsorbates are sensitive to coverage \cite{hertl2026mode} and further detailed studies at higher coverages as well as a comparison with experimental observables, e.g. as previously conducted on ruthenium surfaces~\cite{frischkorn2008ultra}, will be needed to show whether frictional steering and nonadiabatic energy transfer differ in high-coverage scenarios. 
As the rate of laser-assisted diffusion is a key indicator of the magnitude of friction, more experimental studies on diffusion rates, such as via Helium spin echo experiments \cite{rittmeyer2016energ, mcintosh2013quant}  would be of great benefit in the future.

\begin{acknowledgments}
The authors acknowledge support via a UKRI Future Leaders Fellowship [MR/S016023/1], a UKRI Frontier grant [EP/X014088/1], and a Leverhulme Trust Research Project grant [RPG-2019-078] (R.J.M.)
Computational resources were provided by the EPSRC-funded HEC Materials Chemistry (EP/L000202/1, EP/R029431/1) and the HPC-CONEXS (EP/X035514/1) consortia for access to the ARCHER2 UK National Computing Service, and the
EPSRC-funded HPC Midlands+ computing centre for access to Sulis (EP/P020232/1).
\end{acknowledgments}

\paragraph{Data availability}
The software packages used in this work are described in the supporting material and are publicly available, with the exception of FHI-AIMS, which is subject to licensing conditions. 
Processed simulation data used to generate figures is available from Ref. \cite{spears_2026_23079231}. 

\bibliography{references.bib}
\end{document}

\title{Supporting Information for \\ "Role of anisotropic electronic friction in laser-driven hydrogen recombination on copper"}

\author{Alexander Spears}
\affiliation{Department of Chemistry, University of Warwick, Gibbet Hill Road, CV4 7AL Coventry, UK}
\affiliation{University of Vienna, Faculty of Physics, Kolingasse 14-16, Vienna, Austria}
\author{Wojciech G. Stark}
\affiliation{Department of Chemistry, University of Warwick, Gibbet Hill Road, CV4 7AL Coventry, UK}
\affiliation{Department of Chemistry, Imperial College London, White City Campus, Molecular Sciences Research Hub, 82 Wood Ln, London W12 0BZ, UK}

\author{Reinhard J. Maurer}
\email{reinhard.maurer@univie.ac.at}
\affiliation{Department of Chemistry, University of Warwick, Gibbet Hill Road, CV4 7AL Coventry, UK}
\affiliation{University of Vienna, Faculty of Physics, Kolingasse 14-16, Vienna, Austria}
\affiliation{Department of Physics, University of Warwick, Gibbet Hill Road, CV4 7AL Coventry, UK}

%\date{}

\maketitle

\tableofcontents
\clearpage

\section{Computational Details}\label{computational-details}

\subsection{Density functional theory}\label{sec-methods-dft}

Density functional theory calculations were performed using the all-electron numerical atomic orbital code FHI-AIMS \cite{blum2009initi} (version \texttt{240920}) using the SRP48 exchange correlation functional \cite{srp48functional} due to its accuracy in describing the H/Cu system.
This functional contains a 52~\% contribution from the Perdew-Burke-Ernzerhof (PBE) functional \cite{pbefunctional} and a 48~\% contribution from the revised PBE (RPBE) functional \cite{rpbefunctional}.
SCF convergence criteria were set to 10\textsuperscript{-6}~eV for total energies, 10\textsuperscript{-3}~eV for eigenvalue energies, 10\textsuperscript{-5}~e~a\textsubscript{0}\textsuperscript{-3} for charge densities and 10\textsuperscript{-4}~eV~\AA\textsuperscript{-1} for forces.
A $k$-grid of ($12\times12\times 1$) was used for all periodic structures.
All settings are chosen to be consistent with previous calculations with which the machine learning potentials were constructed \cite{stark2024bench}.

Total energies and forces were calculated with the default ``tight'' basis set with additional ``first-tier'' basis functions for copper.
The electron density for different surface configurations used to train LDFA models was calculated using the \texttt{output\_cube total\_density} keyword.

To determine the electronic friction tensor (EFT) from first principles electron phonon coupling (orbital dependent friction, ODF), a ``light'' basis set with additional ``second-tier'' improvements for hydrogen was found to yield sufficiently converged results.
The electronic friction tensor was output using the \texttt{output\_friction\_gamma} keyword.
All ODF evaluation settings are the same as presented in Refs.~\onlinecite{sachs2025machi, stark2025nonad, stark2023machi}.

\subsection{MACE machine learning potential}\label{sec-methods-pes}

MACE \cite{batatia2022mace} models were trained by Stark \textit{et al} using a structure database previously generated to model molecular scattering experiments.
Here, we used the models in ref. \onlinecite{stark2024bench} without further modification as further validation tests confirmed that they are sufficiently accurate to model light-driven recombinative desorption.
In the following, we briefly summarise the training data and model hyperparameters defined in previous work \cite{stark2024bench}.

The cutoff distance for interactions is 5.0~\AA, with a correlation order of 2.
Message passing was set to 16~invariant messages, with an interaction order of 2 and no equivariant message passing.
Model parameters were optimised over 1500~iterations to minimise errors for forces and total energy with two error weighting schemes.
During the first 1200~iterations, force and energy errors were weighted at 95:5.
In the final 300~iterations, total energy errors were weighted at 1000:1 over force errors.

The training data in ref.~\onlinecite{stark2024bench} contains ($3\times 3\times 6$) supercells with $(100), (110), (111)$ surface facets, as well as ($3\times 1\times 6$) supercells with $(211)$ surface facets from \textit{ab-initio} molecular dynamics (AIMD) simulations.
These included simulations of clean Cu surfaces near thermal equilibrium at 300, 600 and 900~K, as well as simulations with chemisorbed H on Cu at 300~K with the bottom two layers of Cu atoms fixed to prevent unit translations.
Further snapshots of molecular scattering processes were included from simulations on ($2\times 2\times 4$) supercells.
Multiple structures with H\textsubscript{2} at larger distances to the surface atoms, where interactions are minimal, were included to sample different bond lengths of H\textsubscript{2} in the gas phase.
Through adaptive sampling, 1700 additional snapshots from scattering simulations using PaiNN \cite{schutt2019unify} and SchNet \cite{schutt2018schne} models were added to the training database for the MACE model, with a total of 4230~structures.

\subsection{Predicting electronic friction
tensors}\label{sec-methods-eft}

LDFA friction models from ref. \onlinecite{stark2025nonad} were used in this work without further adaptation.
These models are based on the Atomic Cluster Expansion using the  \href{https://github.com/ACEsuit/ACEpotentials.jl/}{\texttt{ACEpotentials.jl}} \cite{witt2023acepo} Julia package (version \texttt{0.6.7}).
The expansion was terminated above a correlation order of 3, with maximum polynomial degrees of 12,\ 10,\ and 8.
The cutoff distance was set at 4.0~\AA.

The \href{https://github.com/acesuit/ACEfriction.jl}{\texttt{ACEfriction.jl}} \cite{sachs2025machi} models used in ref. \onlinecite{stark2025nonad} were used in this work without further adaptation.
The atomic cluster expansion was terminated at a correlation order of 2, with 6\textsuperscript{th} order radial basis functions. A cutoff radius for interactions was specified as 5.0~\AA.
The learning rate was set to 10\textsuperscript{-4}~~\cite{stark2025nonad}.

\subsection{Two-temperature model}\label{two-temperature-model}

\begin{figure}
\centering{
\includegraphics[width=0.5\linewidth,height=\textheight,keepaspectratio]{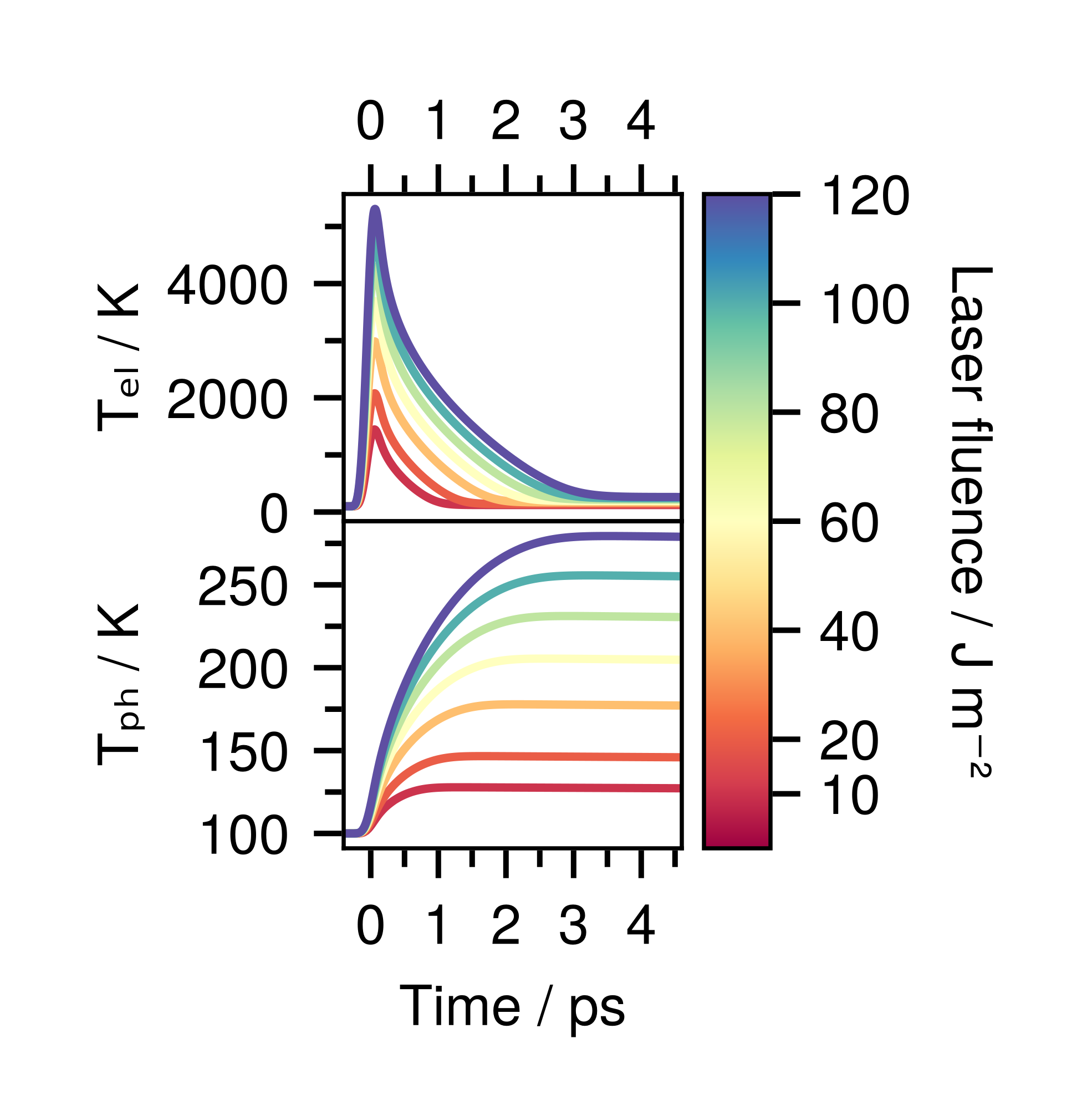}
}

\caption{\label{fig-ttm-progressions}Temperature progressions in the
two-temperature model for bulk Cu at different absorbed laser fluences.}

\end{figure}

A two-temperature model (TTM) was used to model light-matter interactions. Further details about the method are explained in section \ref{sec-explain-TTM}.
All parameters for the TTM were taken from ref.~\onlinecite{scholz2019vibra}. Their values are specified in table \ref{tbl-2TM-parameters}, leading to the temperature progressions shown in figure \ref{fig-ttm-progressions}.

\begin{table*}

\caption{\label{tbl-2TM-parameters}Material parameters for bulk Cu used in the two-temperature model.}

\centering{

    \begin{tabular}{lll} \hline\hline
        Parameter & Explanation & Value in this work\\\hline
        \(g\) & Electron-phonon coupling constant & $1\times 10^{-17}~\mathrm{W~m^{-3}~K^{-1}}$ \\
        \(\xi\) & Laser penetration depth for \(\lambda=400~\mathrm{nm}\) & \(\frac{1}{\xi}=14.9~\mathrm{nm}\) \\
        \(\gamma_{el}\) & Scaling constant for the electronic heat capacity \(C_{el}\). & \(98.0~\mathrm{J~m^{-3}~K^{-2}}\) \\
        \(\theta_D\) & Debye temperature of Cu. & \(343~\mathrm{K}\) \\
        \(N\) & Atom density of bulk Cu & \(8.5\times 10^{28}\) \\
        \(\kappa_{\mathrm{RT}}\) & Thermal conductivity at room temperature & \(401.0~\mathrm{W^{-1}~K^{-1}}\) \\
        \(\mathrm{FWHM}\) & Full width half maximum of the laser pulse &    \(150~\mathrm{fs}\) \\ \hline\hline
    \end{tabular}

}

\end{table*}%

\subsection{Molecular Dynamics simulations}\label{sec-methods-md}
All molecular dynamics simulations in this work were performed using \href{https://github.com/NQCD/NQCDynamics.jl}{\texttt{NQCDynamics.jl}} (version \texttt{0.15.1}) and related Julia packages \cite{gardner2022nqcdz}.
Some structure manipulations were performed using the atomic simulation environment \cite{software-ase}.

\subsubsection{Initial conditions}\label{sec-methods-MonteCarlo}

Initial conditions were generated by Metropolis-Hastings Monte Carlo
sampling of atomic positions. All surface structures were generated
using a temperature-dependent lattice constant determined from a linear
fit for lattice constants measured from NPT dynamics at different
temperatures \cite{stark2023machi}.
For a unit cell size of \(3\times 3\times 6\) layers, structures with
two adsorbed hydrogen atoms were generated with the atoms located in the
hollow adsorption sites. For each temperature, an evenly spaced sample
of 100,000 structures was taken from a single random walk, discarding the
first 1,000 steps.
The initial velocities were generated from a Maxwell-Boltzmann
distribution for the respective temperature. Since the two lowest layers
of the Cu(111) surface are frozen, the initial velocities of the Cu
atoms in these layers were fixed at zero.
MD simulations combining the TTM with electronic friction methods were
initialised 0.4~ps prior to the laser pulse maximum with absorbed
ﬂuences of 10,~20,~40,~60,~80,~100,~and~120~J~m\textsuperscript{-2}. The
intensity maximum of the laser pulse corresponds to \(t=0\). The
electronic friction tensor was calculated using one of the two models
described in section \ref{sec-methods-eft}. Initial positions and
velocities corresponding to thermal equilibrium at
100,~150,~200,~250,~300~K were randomly sampled from the distributions
described above for each trajectory.
TTM-MDEF trajectories were simulated with a time step of 0.1~fs for 5.0~ps
total. Simulations are terminated earlier if the centre of mass of the
H\textsubscript{2} molecule reaches a distance of more than 6~\AA{} from the
surface, indicating desorption.
Table~\ref{tbl-desorption-trj-numbers} gives an overview of the total number of trajectories simulated for each set of initial conditions.

\begin{table*}[htbp]
	\centering
	\caption{Number of total trajectories simulated for a given set of initial conditions and number of desorption events detected. Simulations were performed with an initial temperature of 100~K.}
	\begin{tabular}{cccc} \hline\hline
	  \textbf{Electronic friction approximation} & \textbf{Laser Fluence / J~m\textsuperscript{-2}} & \textbf{Trajectories} & \textbf{Desorptions} \\\hline
	  LDFA & 10 & 2815 & 0 \\
	  LDFA & 20 & 6907 & 0 \\
	  LDFA & 40 & 6853 & 17 \\
	  LDFA & 60 & 8730 & 92 \\
	  LDFA & 80 & 12050 & 360 \\
	  LDFA & 100 & 12276 & 693 \\
	  LDFA & 120 & 12348 & 1150 \\
	  ODF & 10 & 3501 & 0 \\
	  ODF & 20 & 6865 & 0 \\
	  ODF & 40 & 7027 & 0 \\
	  ODF & 60 & 12365 & 8 \\
	  ODF & 80 & 14615 & 16 \\
	  ODF & 100 & 14687 & 56 \\
	  ODF & 120 & 13608 & 125 \\ \hline\hline
	\end{tabular}
	\label{tbl-desorption-trj-numbers}
\end{table*}

\clearpage
\section{Method details}\label{method-details}

\subsection{Two-temperature model}\label{sec-explain-TTM}

The two-temperature model (TTM) is a system of two coupled differential
equations that describe the time evolution of the electronic and
phononic temperatures, \(\mathrm{T_{el}}\) and \(\mathrm{T_{ph}}\). In
this work, we neglect lateral heat diffusion across the metal surface,
allowing for a simplified one-dimensional representation of the
two-temperature model. The TTM is given by the following set of
equations:
\begin{equation}\protect\phantomsection\label{eq-2TM-el}{
\mathrm{C_{el}}\frac{\partial}{\partial t}\mathrm{T_{el}}=\nabla(\kappa\nabla \mathrm{T_{el}})-\mathrm{g(T_{el}-T_{ph})+S}(z,t)
}\end{equation}

\begin{equation}\protect\phantomsection\label{eq-2TM-ph}{
\mathrm{C_{ph}}\frac{\partial}{\partial t}\mathrm{T_{ph}=g(T_{el}-T_{ph})}
}\end{equation}

\begin{equation}\protect\phantomsection\label{eq-2TM-laser}{
\mathrm S(z,t)=\mathrm F\cdot\frac{\exp{\frac{-z}{\xi}}\cdot\exp{\frac{-(t-t_0)^2}{2\tau^2}}}{\xi\cdot\sqrt{2\pi}\tau};
\tau=\frac{\mathrm{FWHM}}{2\sqrt{2\ln{2}}}
}\end{equation}

The electronic heat capacity \(\mathrm{C_{el}(T_{el})}\) uses a linear
scaling relation:

\[
\mathrm{C_{el}(T_{el})}=\gamma_{\mathrm{el}}\mathrm{T_{el}}
\]

The temperature dependence of the thermal conductivity of Cu is modelled
using the following relation:

\[
\kappa(\mathrm{T_{el}, T_{ph}})=\kappa_{\mathrm{RT}}\mathrm{\frac{T_{el}}{T_{ph}}}
\]

The phononic heat capacity is calculated using the Debye model with
\(N\) and \(\theta_D\) values given above:

\[
\mathrm{C_{ph}(T_{ph})}=9N\mathrm{k_B}\left(\frac{\mathrm{T_{ph}}}{\theta_\mathrm D}\right)^3\int_0^{\frac{\theta_{\mathrm{D}}}{\mathrm{T_{ph}}}}\frac{x^4e^x}{(e^x-1)^2}~dx
\]

\subsection{Local Density Friction Approximation
(LDFA) \label{sec-explain-LDFA}}

The local density friction approximation (LDFA) is used to calculate the EFT by determining a spatially isotropic electron density using the free electron gas model \cite{juaristi2008role}.
Each nucleus is treated independently, assuming it to be an ion travelling through a homogeneous electron gas.

The electronic friction coefficient then becomes a function of the electron density $\rho$, Fermi level scattering phase shifts $\delta^{F}$ and the Fermi momentum $k_F$, as shown in eq. \ref{eq-ldfa-EFT}.
\begin{equation}\protect\phantomsection\label{eq-ldfa-EFT}{
\gamma^{\text {LDFA }}=\frac{4 \pi \rho}{\mathrm{k_{F}}} \sum_{l}(l+1) \sin^2 \left(\delta_{1}^{\mathrm F}-\delta_{l+1}^{\mathrm F}\right)
}\end{equation}

Using data from Gerrits \emph{et al.} \cite{gerrits2020elecu}, the electronic friction coefficients as a function of the Wigner-Seitz radius are interpolated
across a range of electron densities using spline fitting to the relation in \cref{eq-ldfa-WignerSeitz}.

\begin{equation}\protect\phantomsection\label{eq-ldfa-WignerSeitz}{
r_s = \left(\frac{3}{4\pi\rho(\mathbf{r_i})}\right)^\frac{1}{3}
}\end{equation}

\subsection{Orbital-Dependent Friction \label{sec-explain-ODF}}

The Orbital-dependent friction (ODF) approach yields a spatially anisotropic EFT, which includes contributions by the electronic structure of adsorbates in the framework of first-order time-dependent perturbation theory within Density Functional Theory \cite{head-gordon1995molec, maurer2016initi}.
In the following, we use the expression by Box \emph{et al} \cite{box2023initi}, which is valid across a range of temperatures and constrain the phonon momentum to \(\mathbf{q}=0\).
For the nonadiabatic coupling between single-particle Kohn-Sham states \(m\) and \(n\), with a wave vector \(\mathbf{k}\), the phonon momentum \(\mathbf{q}\) and spin label \(\sigma\), the EFT can be calculated as follows:

\begin{equation}\protect\phantomsection\label{eq-tdpt}{
    \begin{aligned}
    \Lambda_{\mathbf{q},IJ}(\hbar\omega) = & \pi\hbar \sum_{\mathbf{k}, \sigma, m,n}w_{\mathbf{k}} \tilde{g}^{I}_{mn} (\mathbf{k}, \mathbf{q}) [\tilde{g}^{J}_{mn}]^{*} (\mathbf{k}, \mathbf{q}) \times (f_{n \mathbf{k}}-f_{m \mathbf{k}+ \mathbf{q}}) \frac{\delta(\epsilon_{m \mathbf{k}+ \mathbf{q}}-\epsilon_{n \mathbf{k}}-\hbar\omega)}{\epsilon_{m \mathbf{k}+ \mathbf{q}}-\epsilon_{n \mathbf{k}}}
    \end{aligned},
}\end{equation}

with energies of each single-particle state, \(\epsilon_{m\mathbf{k}+\mathbf{q}}\), their occupations \(f_{n\mathbf{k}}\), and electron phonon coupling matrix elements, $\tilde{g}^{J}_{mn}$ .
This expression is evaluated in the quasi-static limit, where \(\hbar\omega \to 0\).

\subsection{Molecular Dynamics with Electronic
Friction}\label{sec-explanation-MDEF}

Molecular dynamics with electronic friction (MDEF) includes non-adiabatic effects through a mean-field friction force describing the average coupling between all states.
Considering non-adiabatic coupling as a weak perturbation to the nuclear system, the nuclear dynamics can be described by a modified Langevin equation:

\begin{equation}\protect\phantomsection\label{eq-mdef}{
    \mathrm m\frac{\partial^2}{\partial t^2}\mathbf R=\nabla V-\mathbf\Lambda \frac{\partial}{\partial t}\mathbf R+\mathcal{W}(\mathrm{T_{el}}(t), \mathbf\Lambda)
}\end{equation}

The electron and phonon temperatures are propagated as described above, yielding temperature profiles (see \cref{fig-ttm-progressions}) that are applied to the MD simulations through the \(\mathcal{W}\) term in \cref{eq-mdef}.

\(\mathrm{T_{el}}\) is applied directly to the adsorbate atoms, while \(\mathrm{T_{ph}}\) is applied to some or all of the surface atoms according to the schemes shown in fig. \ref{fig-thermostatting-strategies}.
Following a comparison of different strategies for Cu(111) (section \ref{sec-phononThermostat-accuracy}), simulations were performed with four-layer phonon thermostatting as this led to the least deviation from the TTM T\textsubscript{ph}(t) profile for an ensemble of simulated trajectories.

\begin{figure}
\includegraphics[width=0.33\linewidth,keepaspectratio]{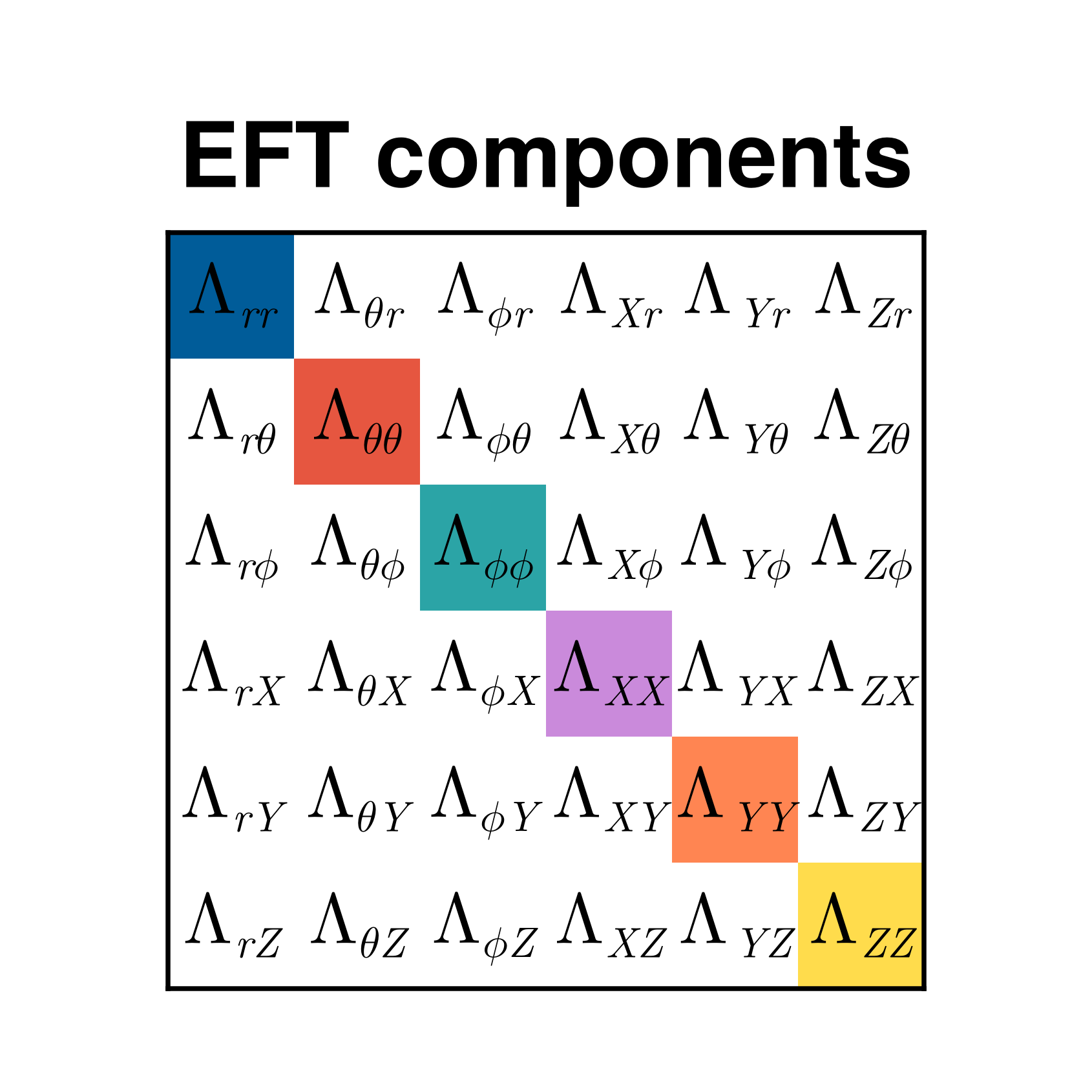}
\hspace{4pt}
\includegraphics[width=0.33\linewidth,keepaspectratio]{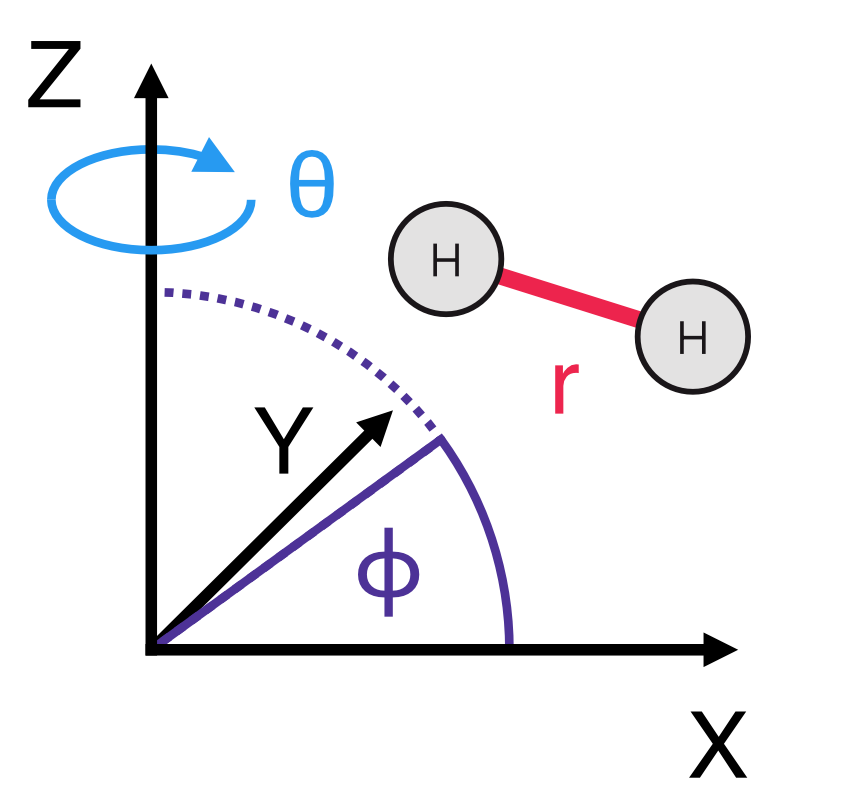}

\caption{\label{fig-EFT-internal-coordinates}Internal coordinate representation of a hydrogen molecule (left) and the corresponding components of the electronic friction tensor (right) described by the coordinate transformation in eq. \ref{eq-EFTtransform}.}

\end{figure}

To quantify mode-specific non-adiabatic coupling, the EFT is transformed into the internal reference frame of a diatomic molecule with the interatomic distance \(r\), polar rotation angles \(\theta\) and \(\phi\), and centre of mass coordinates \(\mathrm{X, Y, Z}\) (see fig. \ref{fig-EFT-internal-coordinates}) using eq. \ref{eq-EFTtransform}.

\begin{equation}\protect\phantomsection\label{eq-EFTtransform}{
    	\Lambda_{\mathrm{int}}=\mathbf{U}^T\mathbf{\Lambda}_{\mathrm{cart}} \mathbf{U}
        }
\end{equation}

The Jacobi matrix \(\mathbf{J}\) is defined as follows:
\begin{equation}\protect\phantomsection\label{eq-jacobiMatrix}{
    \mathbf{J}=\begin{bmatrix}
        \frac{x_1-x_2}{r} & \frac{(x_2-x_1)(z_2-z_1)}{r^2r'} & \frac{y_1-y_2}{r'^2} & \frac{m_1}{m_1+m_2} & 0 & 0\\
        \frac{y_1-y_2}{r} & \frac{(y_2-y_1)(z_2-z_1)}{r^2r'} & \frac{x_2-x_1}{r'^2} & 0 & \frac{m_1}{m_1+m_2} & 0\\
        \frac{z_1-z_2}{r} & \frac{-r'}{r^2} & 0 & 0 & 0 & \frac{m_1}{m_1+m_2}\\
        \frac{x_2-x_1}{r} & \frac{(x_1-x_2)(z_2-z_1)}{r^2r'} & \frac{y_2-y_1}{r'^2} & \frac{m_2}{m_1+m_2} & 0 & 0\\
        \frac{y_2-y_1}{r} & \frac{(y_1-y_2)(z_2-z_1)}{r^2r'} & \frac{x_1-x_2}{r'^2} & 0 &  \frac{m_2}{m_1+m_2} & 0\\
        \frac{z_2-z_1}{r} & \frac{r'}{r^2} & 0 & 0 & 0 & \frac{m_2}{m_1+m_2}\\
    \end{bmatrix}
}\end{equation}

With the distances \(r\) and \(r'\) defined as:

\[
    \begin{aligned}
        r=\sqrt{(x_2-x_1)^2+(y_2-y_1)^2+(z_2-z_1)^2}\\
        r'=\sqrt{(x_2-x_1)^2+(y_2-y_1)^2}\\
    \end{aligned}
\]
To ensure the transformation is trace-preserving, each column of $\mathbf J$ is normalised to a unit vector, yielding $\mathbf U$.
\begin{equation}\label{eq-normalise}
	\mathbf U = \mathbf J \cdot \begin{bmatrix}
		\frac{1}{\sum_i U_{i1}} & 0 & 0 & 0 & 0 & 0\\
		0 & \frac{1}{\sum_i U_{i2}} & 0 & 0 & 0 & 0\\
		0 & 0 & \frac{1}{\sum_i U_{i3}} & 0 & 0 & 0\\
		0 & 0 & 0 & \frac{1}{\sum_i U_{i4}} & 0 & 0\\
		0 & 0 & 0 & 0 & \frac{1}{\sum_i U_{i5}} & 0\\
		0 & 0 & 0 & 0 & 0 & \frac{1}{\sum_i U_{i6}}\\
	\end{bmatrix}
\end{equation}

\subsection{Desorption probabilities}\label{desorption-probabilities}
The desorption probability is determined by the number of simulation trajectories in which H\textsubscript{2} desorption was observed compared to the total number of simulations that were performed.
A bootstrapping approach is used to resample all trajectories 1000 times to determine a mean and a standard deviation of the probability.
\[
\mathrm{P_{des} = \frac{N_{des}}{N_{total}}}
\]

H\textsubscript{2} molecules are counted as desorbed when the centre of mass of the molecule reaches a distance higher than 6.0~\AA~from the average position of the top layer of surface slab Cu atoms.
Due to the cutoff radii of the ML models, re-adsorption is extremely unlikely at this point.
The transition state to desorption is determined based on the mean H-H distance after desorption has begun.
Searching backward in time from the point of desorption for at least 20~fs, the onset of desorption is reached when the H-H bond distance increases beyond one standard deviation of the mean (\(\mu+\sigma\)).

\subsection{Quantifying surface-adsorbate energy
transfer}\label{quantifying-surface-adsorbate-energy-transfer}

We apply the semiclassical Einstein-Brillouin-Keller (EBK) quantisation \cite{larkoski2006numer} method to approximatively determine the vibrational and rotational quantum numbers \(\nu\) and \(J\) respectively, as \href{https://nqcd.github.io/NQCDynamics.jl/stable/initialconditions/ebk/}{implemented
in NQCDynamics.jl} \cite{gardner2022nqcdz} (as of version \texttt{1.0}).

\subsubsection{Energy partitioning across translation, rotation and
vibration}\label{energy-partitioning-across-translation-rotation-and-vibration}

For classical H\textsubscript{2} in the gas phase, we can partition the
total energy into the following contributions: \[
\mathrm{E_{tot} = E_{pot} + E_{trans} + E_{rot} + E_{vib}}
\]

The classical H\textsubscript{2} molecule has six total degrees of
freedom: Three translational, two rotational and one vibrational.

Assuming the molecule to be a rigid rotator vibrating harmonically, we
can calculate the classical energies using the following expressions: \[
\mathrm{E_{kin}} = \frac{m}{2}|\bar{v}|^2
\]

\[
\mathrm{E_{rot}} = \frac{\mathbf{L}^2}{2\mu \mathrm{I}}; \mu = \frac{m_1m_2}{\sum_i m_i}
\]

\[
\mathrm{E_{vib}} = \left(\nu + \frac{1}{2}\right)*\sqrt{\frac{k}{\mu}} ; k=\frac{\partial}{\partial \mathbf{R}}V(\mathbf{R_0})
\]

Various experimental and theoretical investigations on the
H\textsubscript{2}/Ru(0001) system \cite{wagner2005energ, fuchsel2011dissi} have previously compared the partitioning of energy across degrees of freedom in desorbing molecules.
For a set of simulations performed under the same starting conditions, we determine the ratio of energy partitioning by calculating the mean energy per degree of freedom over all simulations.
Mean energies are determined through bootstrapping with 1000 re-sampling steps.

\[
\bar E_\mathrm{trans} : \bar E_\mathrm{rot} : \bar E_\mathrm{vib}
\]

\subsection{Minimum Energy Paths for desorption}

Minimum energy paths for recombinative H\textsubscript{2} desorption from Cu(111) were calculated using the previously described MACE models.
As shown in the supplementary material in ref. \onlinecite{stark2024bench}, these models are sufficiently accurate in reproducing the desorption barriers obtained using density functional theory.
The climbing-image nudged elastic band algorithm implemented in ase \cite{software-ase} was used to optimise the minimum energy path with 25 intermediate images created by linear interpolation of H atom positions between the adsorbed and desorbed state.
H atoms were initially placed in \textit{hollow} sites, with the minimum energy path leading through the \textit{bridge} regions.
The FIRE optimiser was used with a force accuracy criterion of $10^{-2}$~eV~A\textsuperscript{-1}.

\clearpage
\section{Machine learning potential validation for laser-driven
desorption}\label{machine-learning-potential-validation-for-laser-driven-desorption}

\subsection{MACE potential energy
surface}\label{mace-potential-energy-surface}

To validate MACE model performance with reference to DFT, we calculate
reference energies and forces for a number of structures. While ML model
ensemble uncertainty often gives an indication of the error compared to
reference data, this correlation becomes less reliable as
certain thresholds of accuracy are reached. Since surface diffusion is
the most ``out-of-dataset'' condition for the model previously developed for reactive surface scattering, we pick structures from each
desorption trajectory based on the following criteria:

\begin{enumerate}
\def\labelenumi{\arabic{enumi}.}
\item
  30 evenly spaced structures around and after the transition state for
  desorption.
\item
  30 structures with the highest ensemble energy variance.
\item
  30 structures with the lowest ensemble energy variance.
\item
  30 structures with the highest summed hydrogen force variance.
\end{enumerate}

Comparing energies and forces with reference to DFT for the first three
structure sets will enable an assessment of how reliable ensemble
uncertainty is as an error metric for simulations on each surface facet,
while comparing the first and last data sets should give insight as to
whether force variance is indicative of inaccuracy in describing the
motion of H atoms.

\begin{figure}[htbp]
\centering
\includegraphics[width=0.4\linewidth,keepaspectratio]{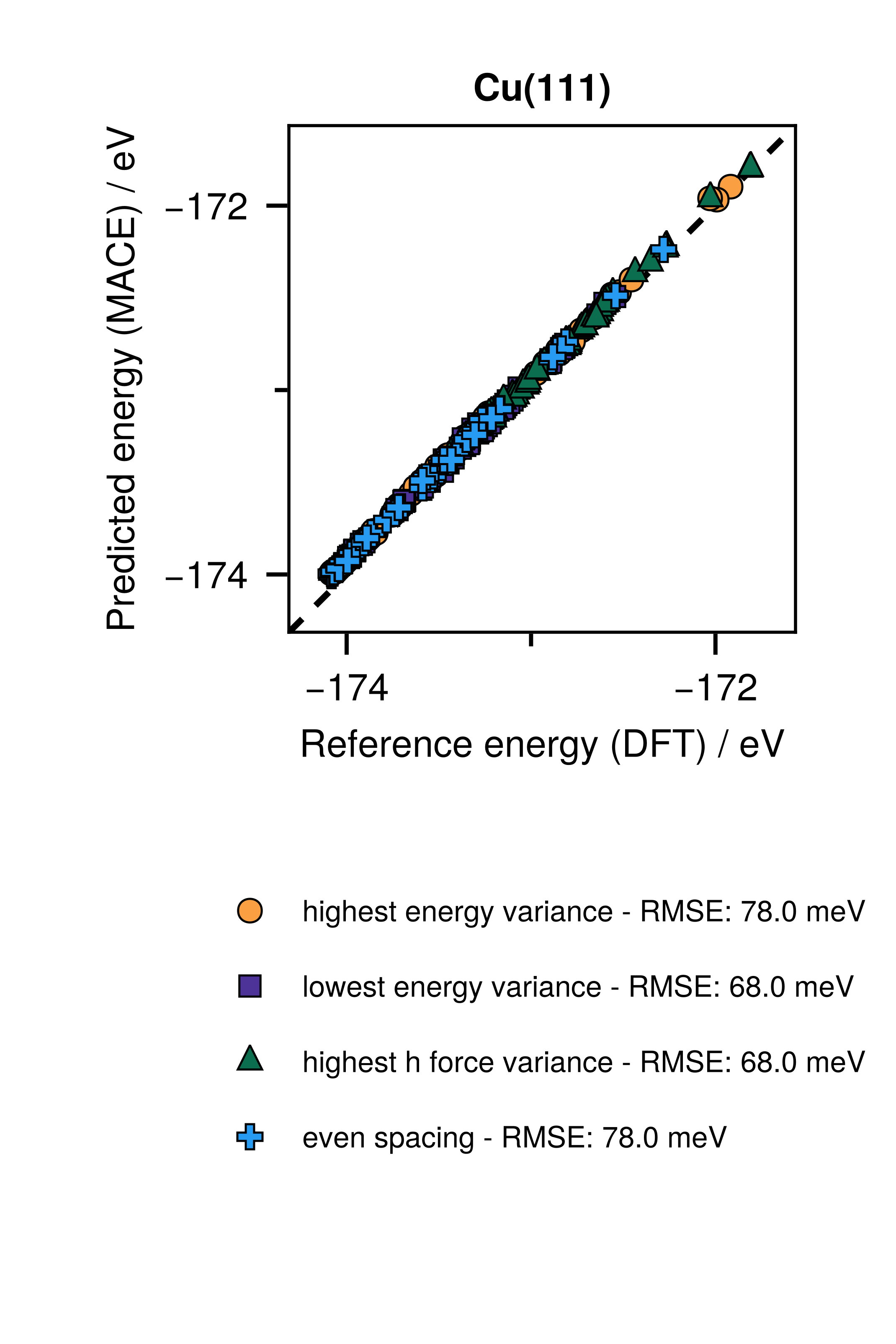}
\hspace{2pt}
\includegraphics[width=0.4\linewidth,keepaspectratio]{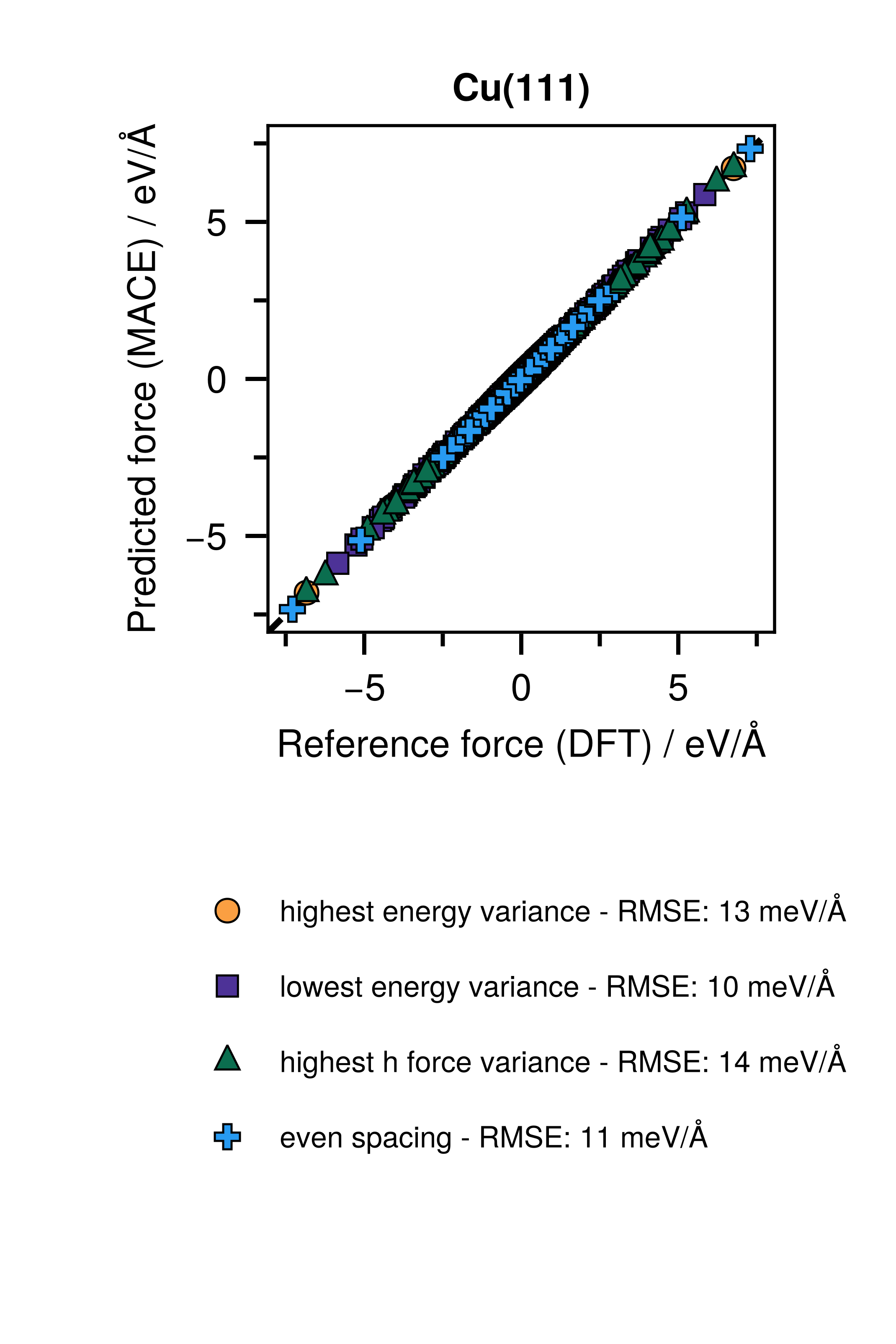}
\caption{\label{fig-correlation-mace-cu111} Energy (left) and Force (right) correlation plots between
MACE model predictions and reference DFT data for a set of Cu(111)
structures. The root mean square error for each set of structures is
shown below.}

\end{figure}%

The RMSEs for evenly spaced and high-variance structures are similar (see fig. \ref{fig-correlation-mace-cu111}),
indicating that there is no significant correlation between ensemble
variance and deviation from reference data.
RMSEs are also similar for structures with the highest hydrogen force variances and evenly sampled structures (see fig. \ref{fig-correlation-mace-cu111}), indicating that the models reproduce hydrogen forces well overall.
As a result, we believe these MACE models are accurate enough to describe laser-driven desorption.

\subsection{ACE-friction ODF models}\label{sec-desorption-eft-values}

\begin{figure}[htbp]
\centering
\includegraphics[width=0.35\linewidth]{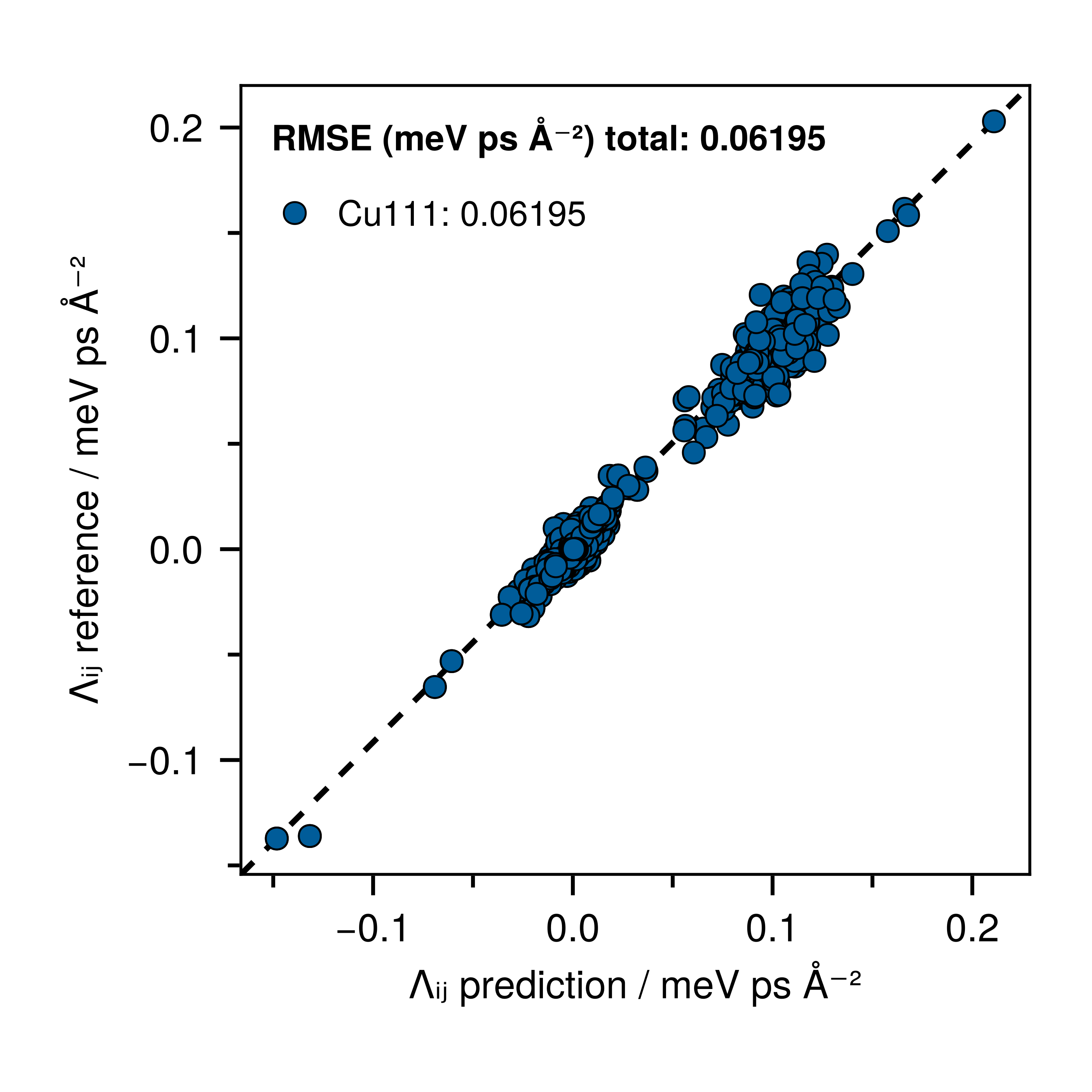}
\hspace{2pt}
\includegraphics[width=0.35\linewidth,keepaspectratio]{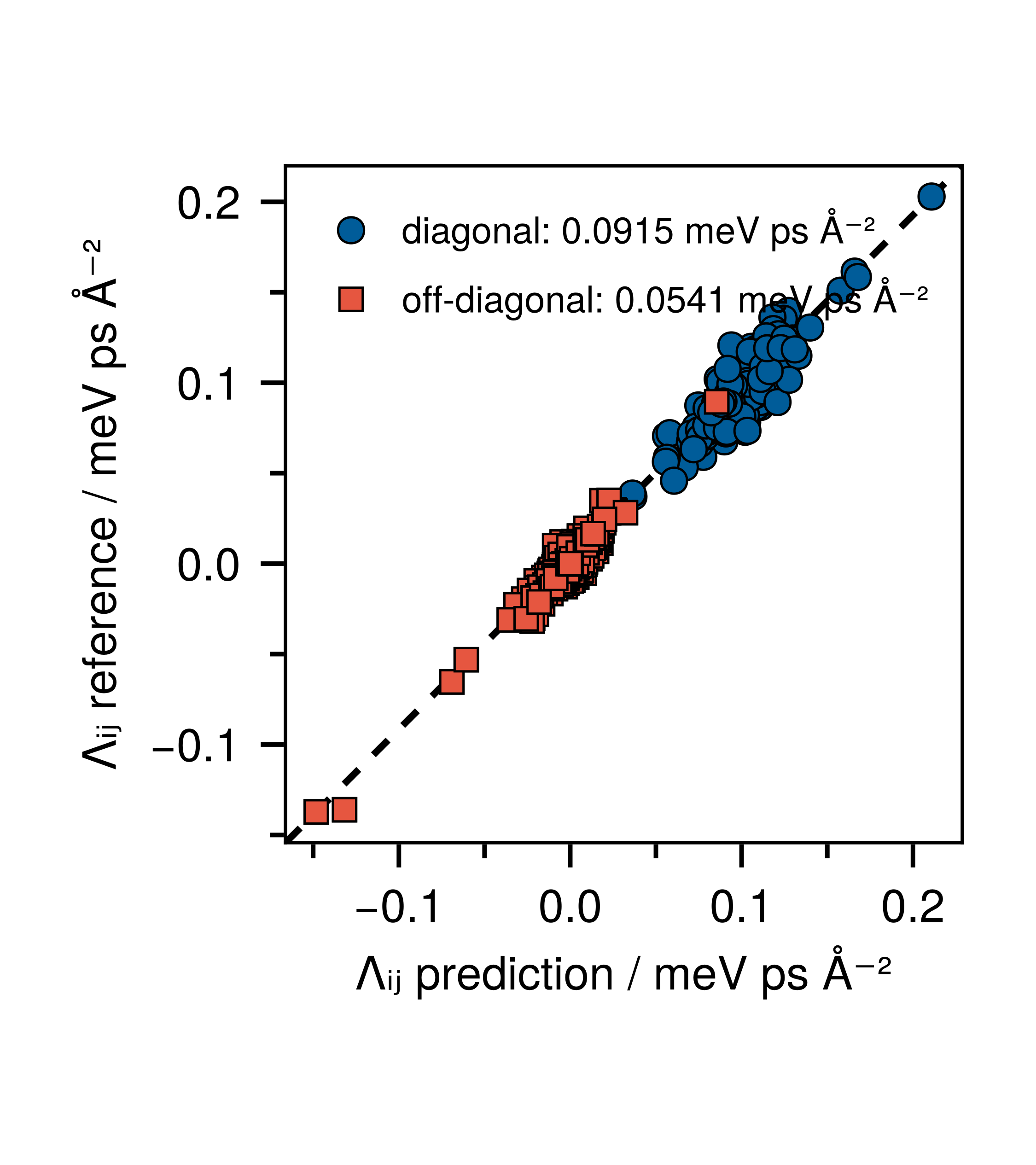}

\caption{\label{fig-MLValidation-EFT-correlation-Cu111}Correlation plots between ACE-friction predicted ODF friction tensors and reference data calculated using FHI-AIMS.}

\end{figure}

\begin{figure}[htbp]

{\centering \includegraphics[width=0.7\linewidth,height=\textheight,keepaspectratio]{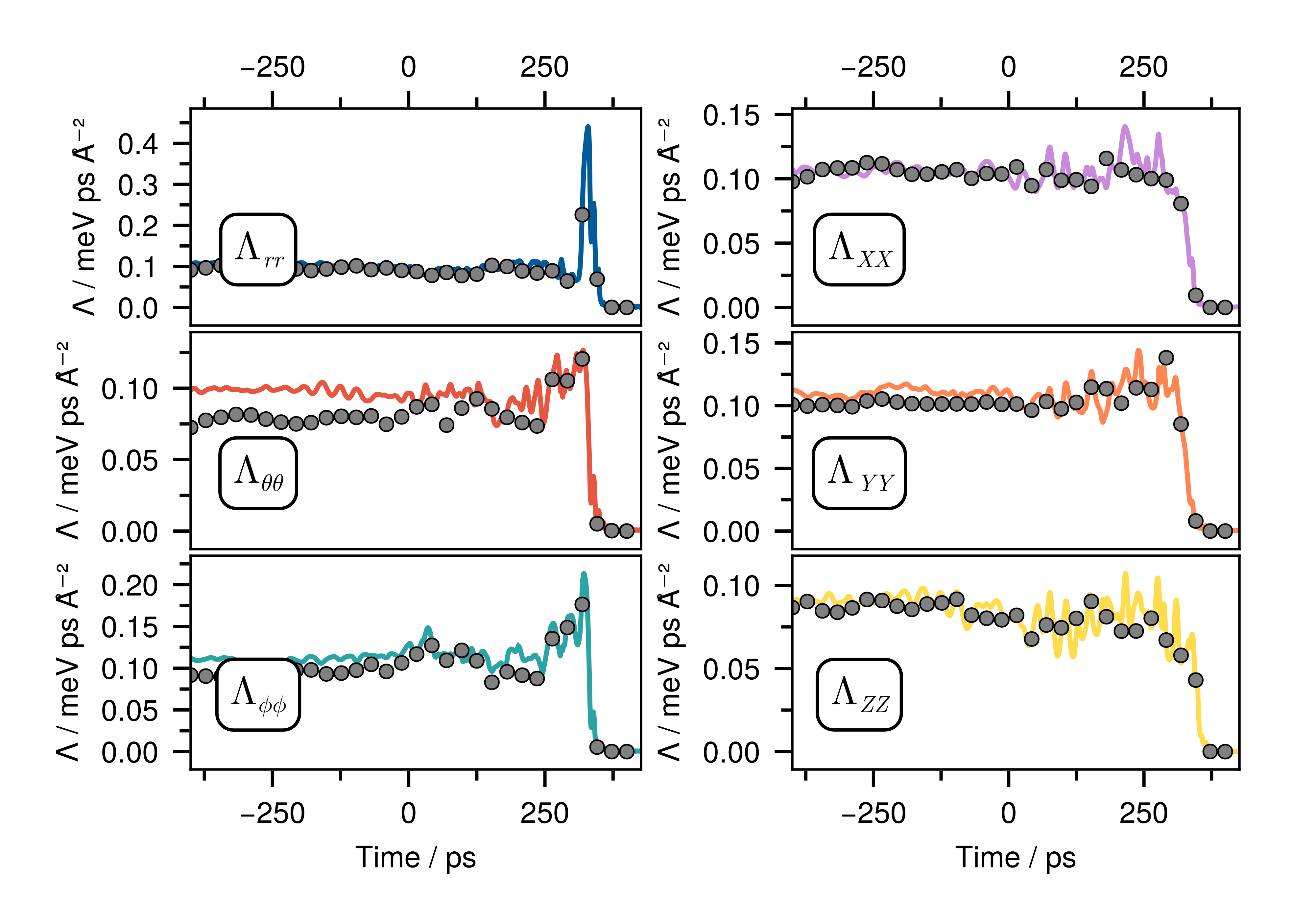}

}

\caption{Component-wise comparison of the accuracy of the predicted EFT
with calculated values for a laser-driven desorption trajectory from
Cu(111). Calculated values are indicated in grey circles. }
\label{fig-eft-components-over-time}

\end{figure}

To assess the accuracy of the \texttt{ACEfriction.jl} \cite{sachs2025machi} model used to predict the ODF electronic friction tensor (EFT) for simulations in this work, we compare predicted EFTs to their reference values calculated using time-dependent perturbation theory (TDPT) for the structures sets previously described.
The DFT settings used to calculate the EFT are described in section \ref{sec-methods-dft}.
For evenly sampled MD snapshots across multiple simulations using ODF, predicted EFT components are in general agreement with their DFT reference values (see \cref{fig-MLValidation-EFT-correlation-Cu111}).
Since the off-diagonal components of the ODF EFT are generally lower in magnitude, the RMSE is slightly lower than for the larger diagonal components.
Along a single trajectory (see fig. \ref{fig-eft-components-over-time}), the predicted EFT components are in line with reference data from ODF EFT calculations.

\clearpage
\section{Impact of phonon
thermostatting}\label{sec-phononThermostat-accuracy}
\begin{figure}[htbp]

\centering{

\includegraphics[width=0.66\linewidth,height=\textheight,keepaspectratio]{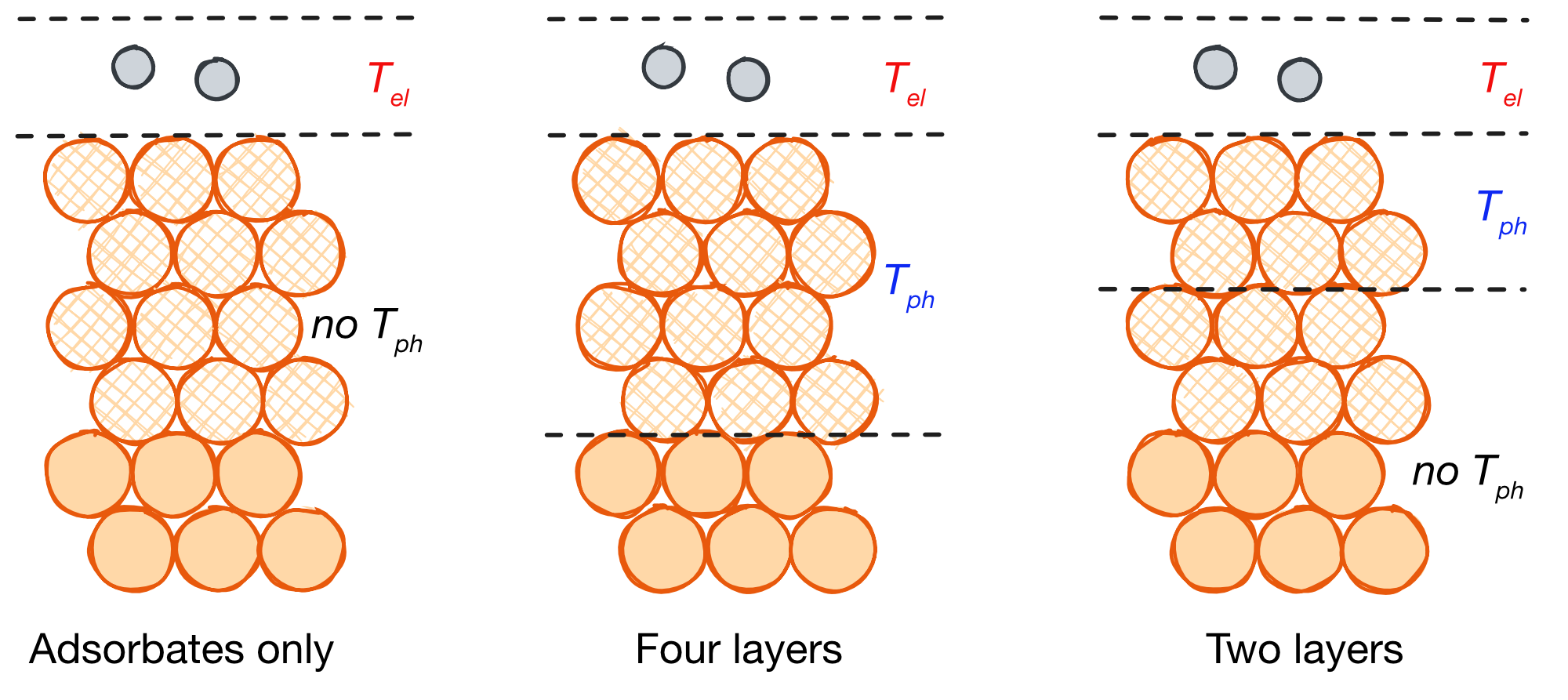}

}

\caption{\label{fig-thermostatting-strategies}Thermostatting strategies
for MDEF + TTM simulations. In this document, desorption probabilities are denoted as \(P_N\), where \(N\) indicates the number of slab layers (counted from the top of the slab) the phonon thermostat is applied to.}

\end{figure}%

\begin{figure}[htbp]
\centering
\includegraphics[width=0.48\linewidth,height=\textheight,keepaspectratio]{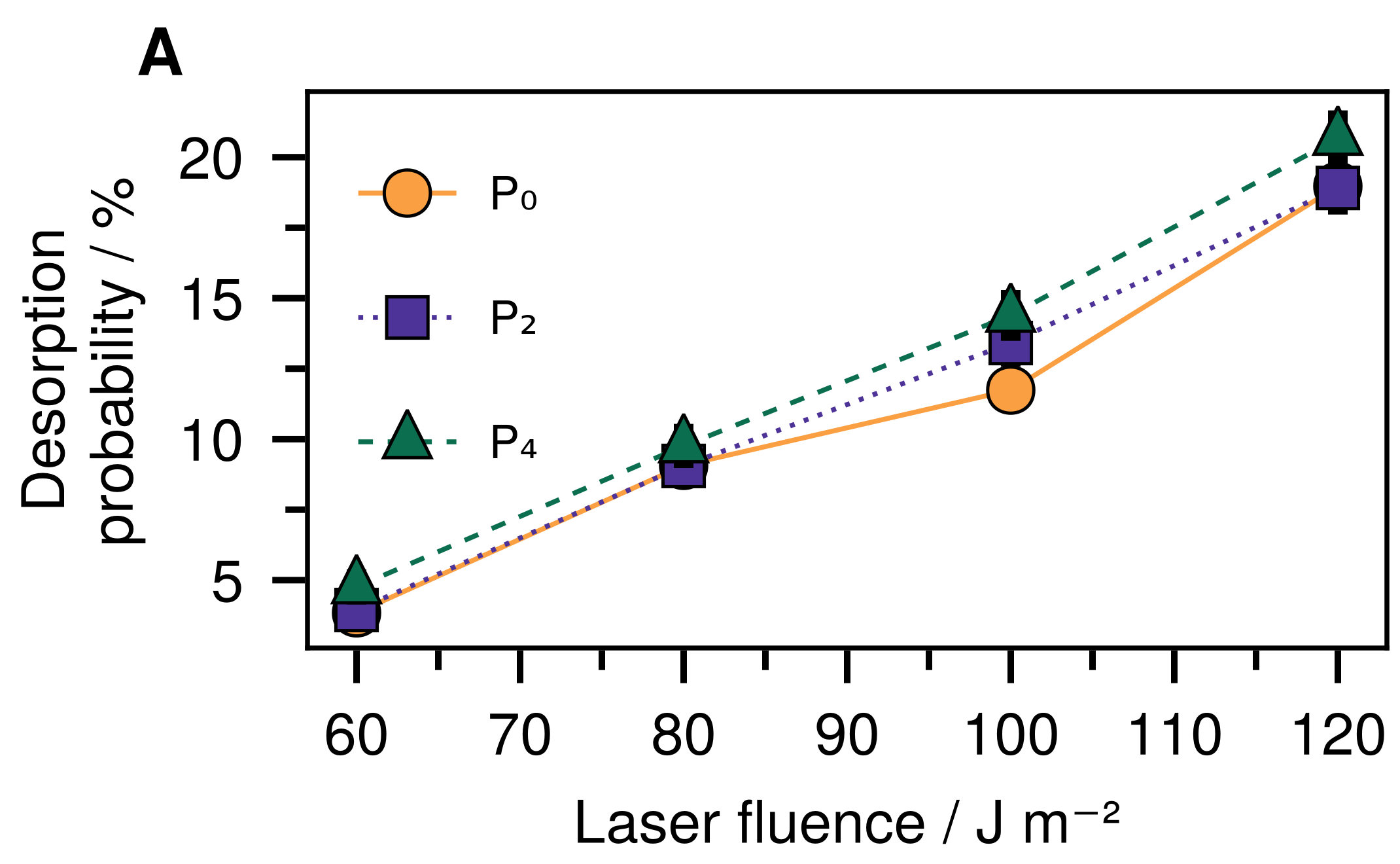}
\hspace{2pt}
\includegraphics[width=0.48\linewidth,height=\textheight,keepaspectratio]{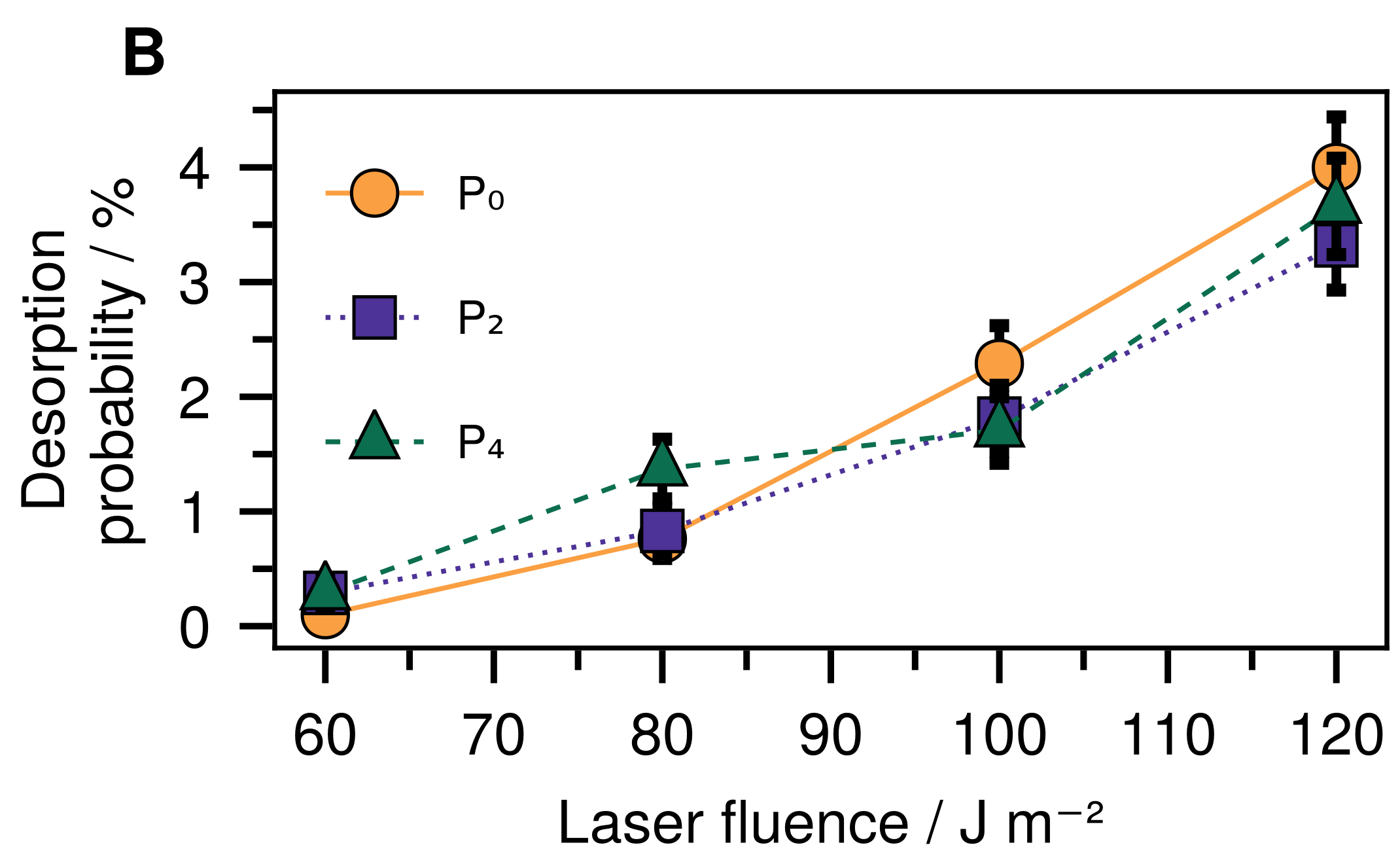}
\caption{\label{fig-DesorptionProbabilityThermostatting}Desorption
probabilities as a function of laser fluence for different thermostatting strategies (see fig. \ref{fig-thermostatting-strategies}) using LDFA (left) and ODF (right) with a starting surface temperature of 200~K. Error bars indicate statistical sampling error based on Bootstrapping the complete set of trajectories. If error bars are not visible, the error is smaller than the markers used.}

\end{figure}

\begin{figure}

\centering{

\includegraphics[keepaspectratio]{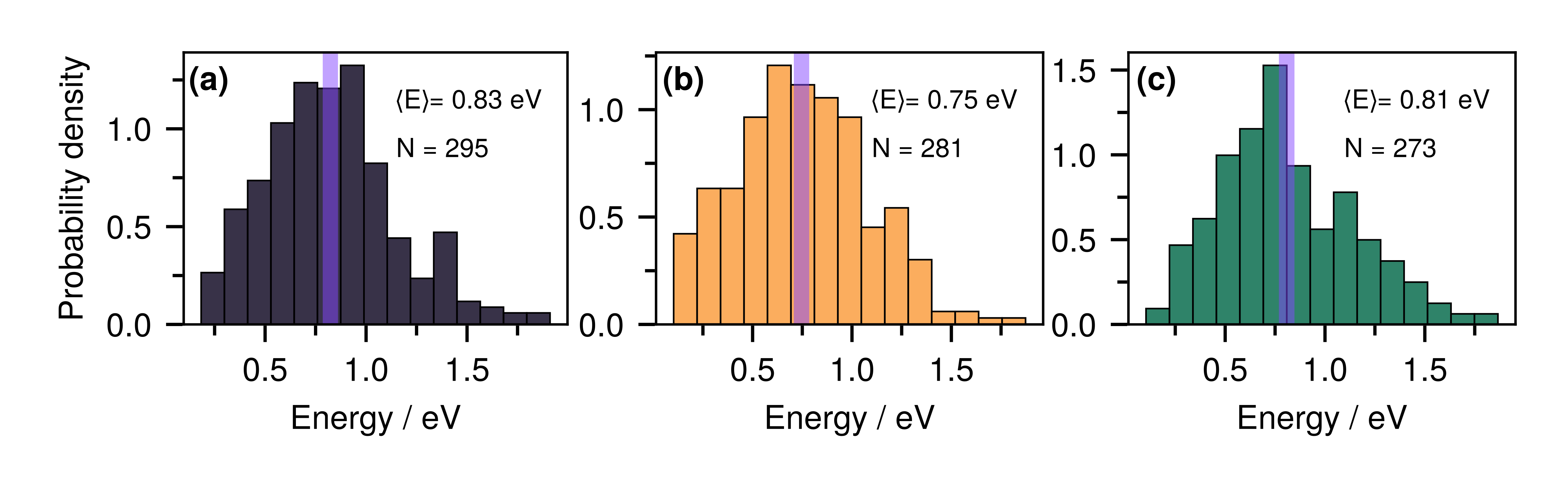}

}

\caption{\label{fig-thermostatting-TranslationEnergy-LDFA}Translational
energy distributions of desorbing H\textsubscript{2} molecules with a
starting temperature of 200~K and an applied laser fluence of
100~J~m\textsuperscript{-2} for LDFA simulations with \textbf{(a):} No
phonon thermostat applied. \textbf{(b)}: Four layers thermostatted.
\textbf{(c)}: Two layers thermostatted. A 95\% confidence interval around
the energy mean is shaded in purple.}

\end{figure}

\begin{figure}

\centering{

\includegraphics[keepaspectratio]{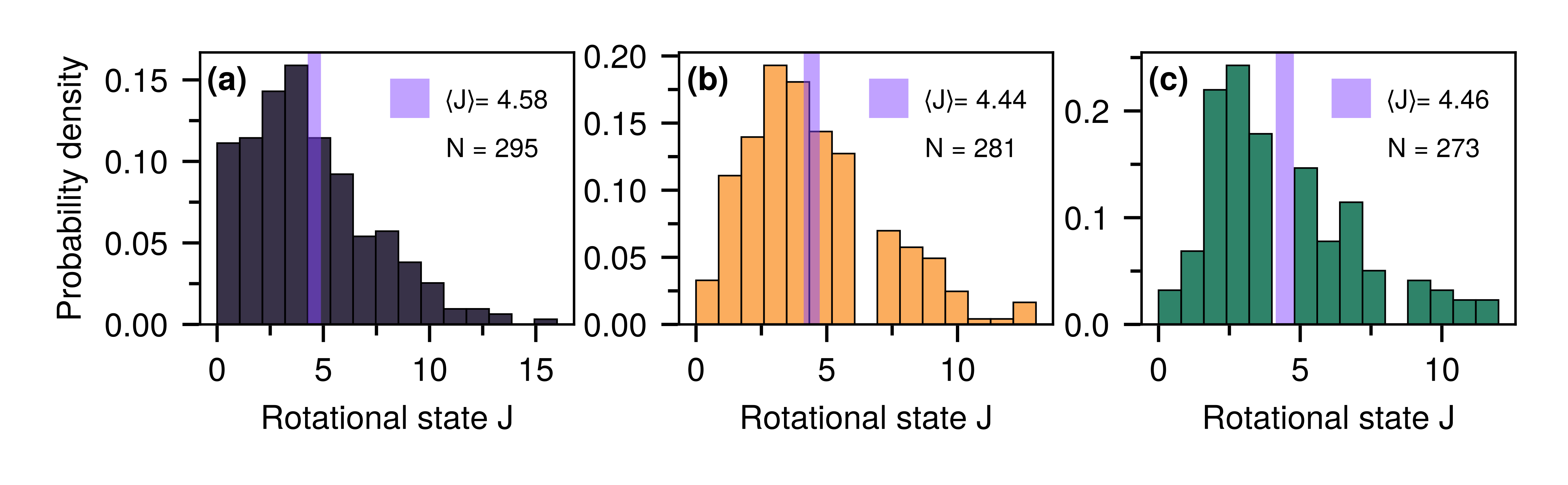}

}

\caption{\label{fig-thermostatting-RotationEnergy-LDFA}Rotational
quantum state distributions of desorbing H\textsubscript{2} molecules
with a starting temperature of 200~K and an applied laser fluence of
100~J~m\textsuperscript{-2} for LDFA simulations with \textbf{(a)}: No
phonon thermostat applied. \textbf{(b)}: Four layers thermostatted.
\textbf{(c)}: Two layers thermostatted. A 95\% confidence interval around
the energy mean is shaded in purple.}

\end{figure}

\begin{figure}

\centering{

\includegraphics[keepaspectratio]{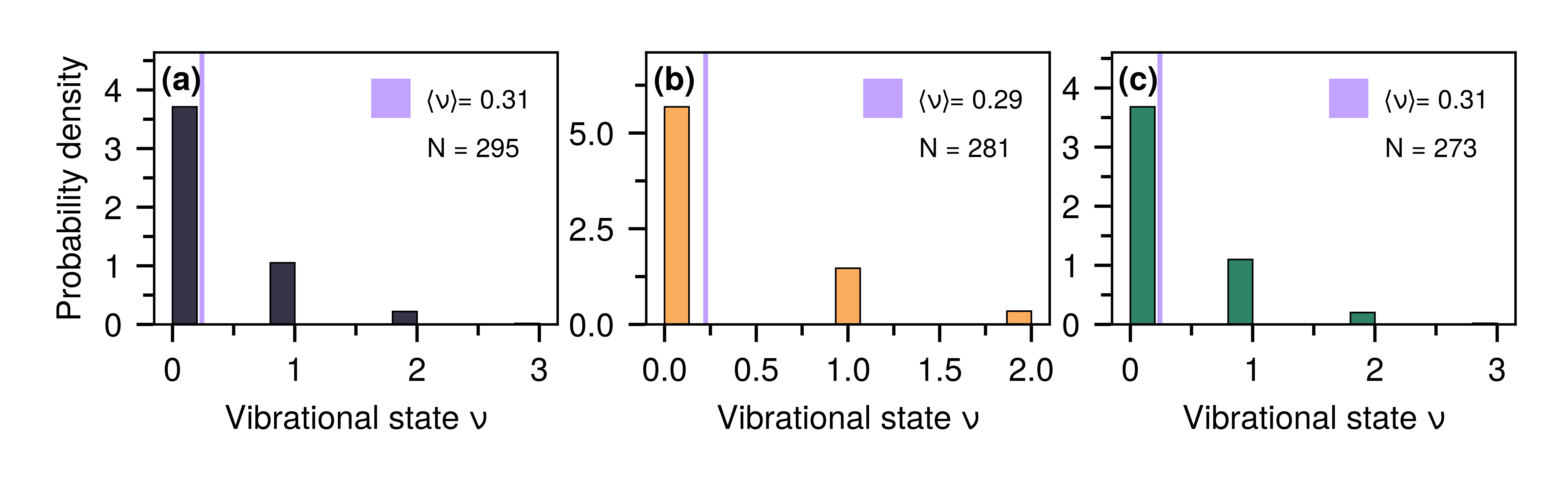}

}

\caption{\label{fig-thermostatting-VibrationEnergy-LDFA}Vibrational
quantum state distributions of desorbing H\textsubscript{2} molecules
with a starting temperature of 200~K and an applied laser fluence of
100~J~m\textsuperscript{-2} for LDFA simulations with \textbf{(a)}: No
phonon thermostat applied. \textbf{(b)}: Four layers thermostatted.
\textbf{(c)}: Two layers thermostatted. A 95\% confidence interval around
the energy mean is shaded in purple.}

\end{figure}

As the choice of phonon thermostatting method (used in
\cref{fig-thermostatting-strategies}) causes additional lattice motion, we investigated the impact of different thermostatting strategies on laser-driven hydrogen recombination for Cu(111).
Overall, the choice of phonon thermostatting scheme does not significantly impact desorption probabilities for LDFA  or ODF  simulations (see fig. \ref{fig-DesorptionProbabilityThermostatting} and tables \ref{tbl-phonons-ldfa}, \ref{tbl-phonons-odf}).
In addition, the energy distributions of desorbing H\textsubscript{2} molecules across translational (\cref{fig-thermostatting-TranslationEnergy-LDFA}), rotational (\cref{fig-thermostatting-RotationEnergy-LDFA}) and vibrational (\cref{fig-thermostatting-VibrationEnergy-LDFA}) degrees of freedom are not significantly affected by the phonon thermostatting approach.

\begin{table*}[htbp]
    \centering
    \caption{Desorption probabilities from a Cu(111) surface for different phonon thermostatting strategies in LDFA simulations. $\mathrm N$ refers to the total number of trajectories simulated for a set of initial conditions. }
    \begin{tabular}{cccccccccc} \hline\hline
        Laser fluence [J~m\textsuperscript{-2}] & P\textsubscript{0} [\%] & Std(P\textsubscript{0}) [\%] & N(P\textsubscript{0}) & P\textsubscript{4} [\%] & Std(P\textsubscript{4}) [\%] & N(P\textsubscript{4}) & P\textsubscript{2} [\%] & Std(P\textsubscript{2}) [\%] & N(P\textsubscript{2}) \\\hline
        60 & 3.84 & 0.35 & 3073 & 4.76 & 0.39 & 2984 & 3.94 & 0.35 & 3073 \\
        80 & 9.07 & 0.50 & 3383 & 9.76 & 0.53 & 2993 & 9.05 & 0.53 & 3073 \\
        100 & 11.74 & 0.52 & 4098 & 14.39 & 0.64 & 3008 & 13.41 & 0.62 & 3073 \\
        120 & 18.97 & 0.71 & 2994 & 20.67 & 0.71 & 3005 & 18.91 & 0.67 & 3073 \\  \hline\hline
    \end{tabular}
    \label{tbl-phonons-ldfa}
\end{table*}

\begin{table*}[htbp]
    \centering
    \caption{Desorption probabilities from a Cu(111) surface for different phonon thermostatting strategies in ODF simulations. $\mathrm N$ refers to the total number of trajectories simulated for a set of initial conditions.}
    \begin{tabular}{cccccccccc} \hline\hline
        Laser fluence [J~m\textsuperscript{-2}] & P\textsubscript{0} [\%] & Std(P\textsubscript{0}) [\%] & N(P\textsubscript{0}) & P\textsubscript{4} [\%] & Std(P\textsubscript{4}) [\%] & N(P\textsubscript{4}) & P\textsubscript{2} [\%] & Std(P\textsubscript{2}) [\%] & N(P\textsubscript{2}) \\\hline
        60 & 0.10 & 0.07 & 2048 & 0.29 & 0.12 & 2048 & 0.29 & 0.11 & 2048 \\
        80 & 0.76 & 0.18 & 2495 & 1.37 & 0.25 & 2048 & 0.83 & 0.20 & 2048 \\
        100 & 2.29 & 0.32 & 2048 & 1.71 & 0.29 & 2048 & 1.81 & 0.29 & 2048 \\
        120 & 4.00 & 0.44 & 2048 & 3.66 & 0.41 & 2048 & 3.32 & 0.39 & 2048 \\  \hline\hline
    \end{tabular}
    \label{tbl-phonons-odf}
\end{table*}

\clearpage
\section{Electronic friction tensor values for example trajectories.}
\begin{figure}[hbtp]
  \includegraphics{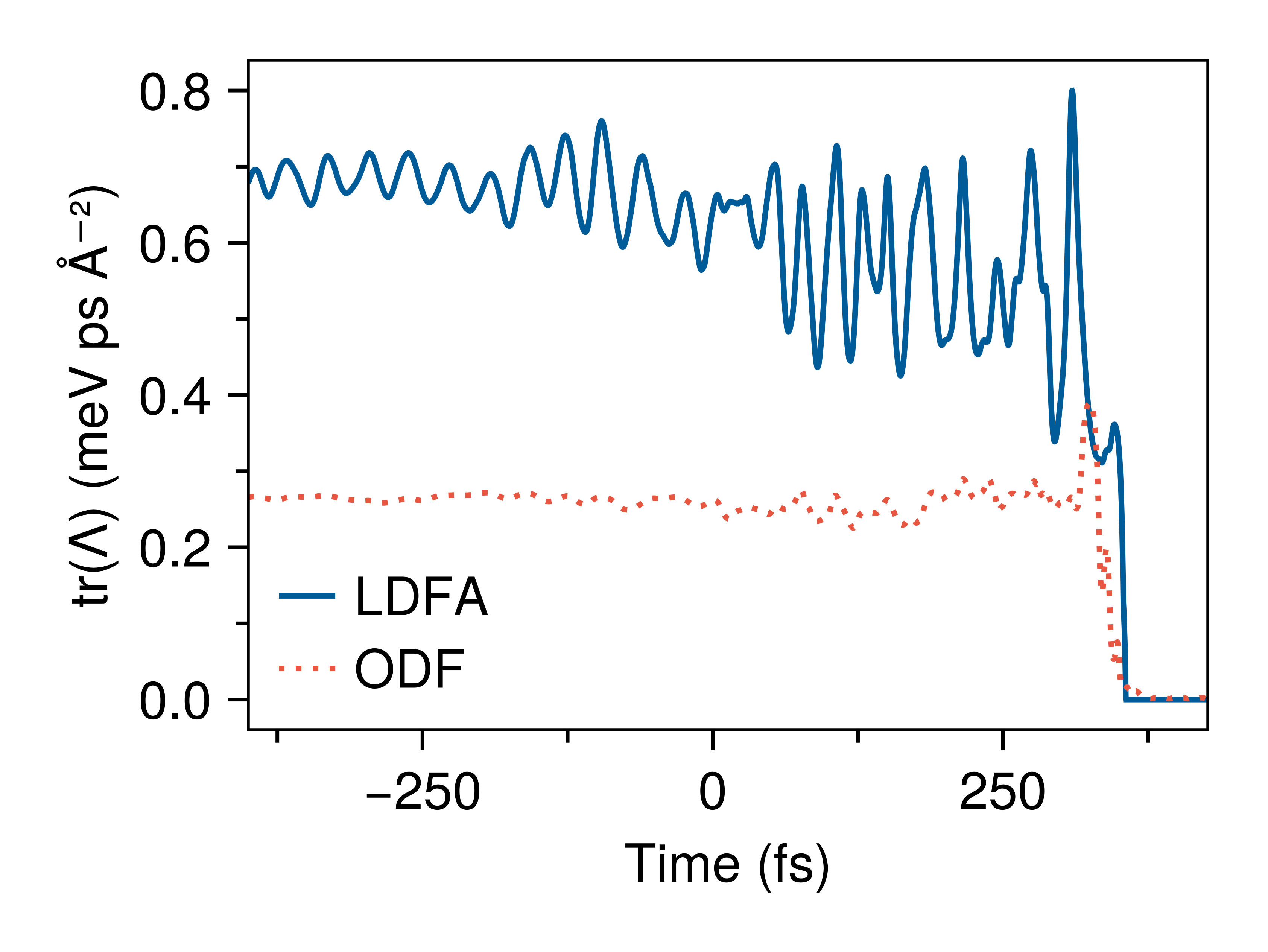}
  \caption{Sum of diagonal elements of the LDFA and ODF electronic friction tensors calculated for an example trajectory. This trajectory was originally simulated using ODF. }
  \label{fig-si-extra01}
\end{figure}

In the following, we provide example values for the electronic friction tensor across LDFA and ODF simulations.
Furthermore, we compare the electronic friction tensor calculated for the same trajectory using LDFA and ODF to deliver a direct comparison of EFT values for typical configurations throughout trajectories.
This enables us to gauge differences between the two electronic friction approximations.
We compare the trace of the EFT for LDFA and ODF calculated for the same trajectory in fig. \ref{fig-si-extra01}.

\begin{figure}[htbp]
  \includegraphics{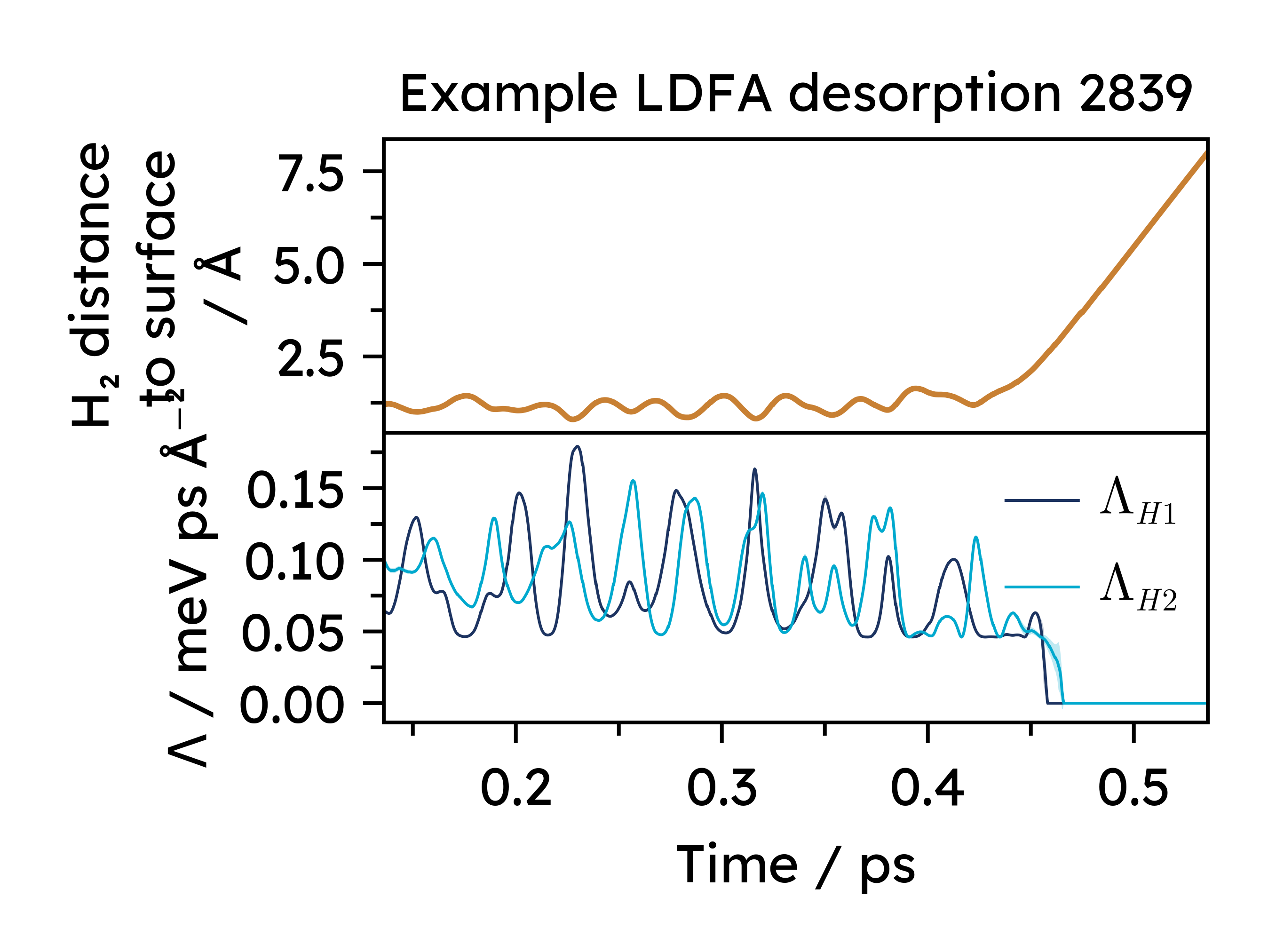}
  \caption{LDFA friction coefficients predicted by ML friction models over a sample desorption trajectory. Ensemble uncertainty of the ML models is indicated with a band around the plotted values. }
\end{figure}

\begin{figure}[htbp]
  \includegraphics[width=0.66\linewidth]{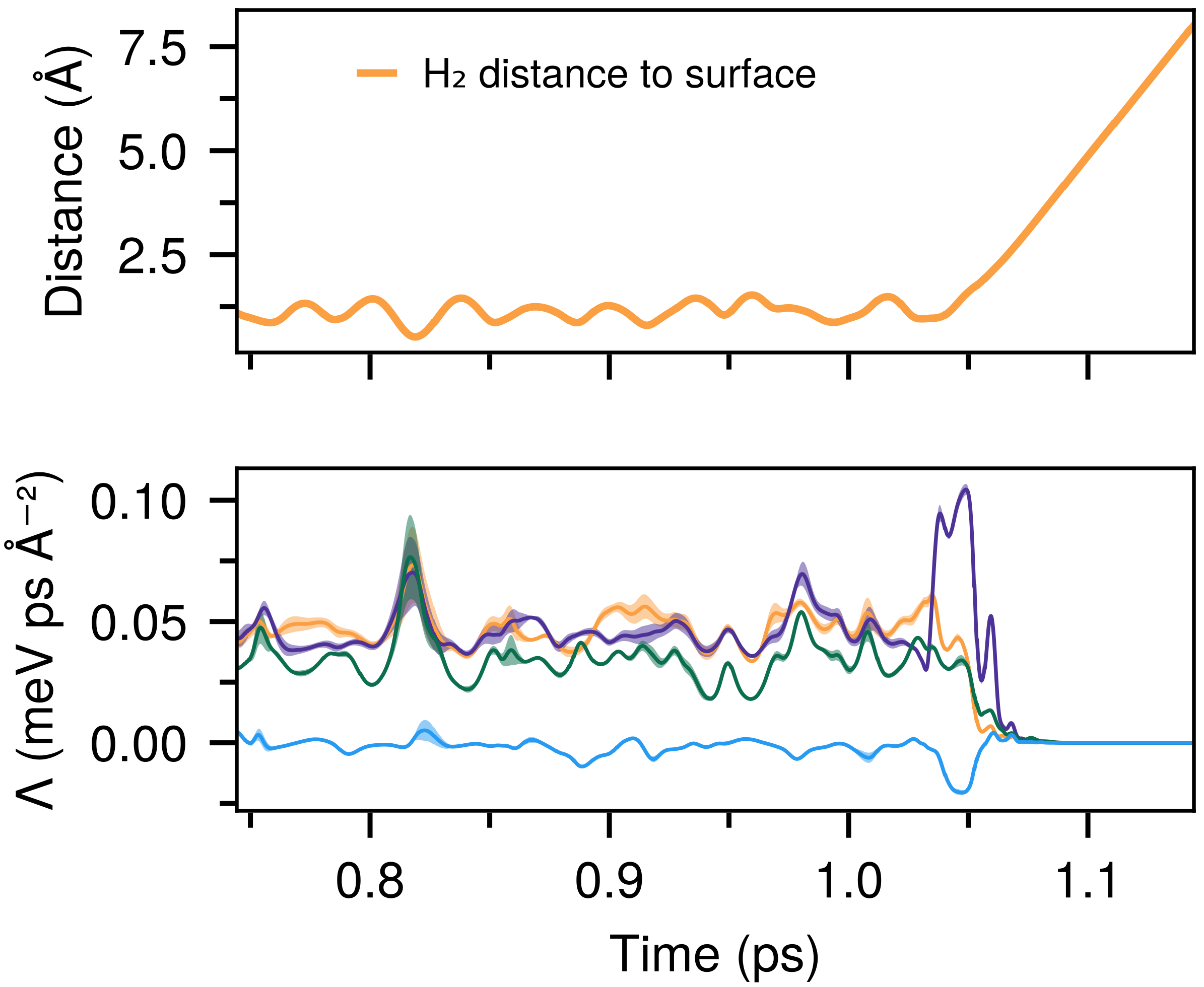}
  \caption{ODF friction coefficients predicted by ML friction models over a sample desorption trajectory. Ensemble uncertainty of the ML models is indicated with a band around the plotted values. \textbf{Orange:} $\Lambda_{x_1x_1}$, \textbf{Purple:} $\Lambda_{y_1y_1}$, \textbf{Green:} $\Lambda_{z_1z_1}$, \textbf{Blue:} $\Lambda_{x_1y_1}$}
\end{figure}

\section{Uncertainty quantification for desorption power law fitting}

The uncertainty of the desorption probabilities propagates into the fitting coefficients.
To quantify this uncertainty, each desorption probability is resampled from the respective distribution of simulations.
These probabilities are then used in nonlinear least squares fitting to equation \ref{eqn-powerlaw}, with a common $\mathrm A$ parameter between LDFA and ODF.

	\begin{equation}\label{eqn-powerlaw}
		\mathrm{Y} = \mathrm{|A|\cdot F^n}
	\end{equation}

By resampling the desorption probabilities 4000 times, we can quantify the uncertainty in the resulting fitting coefficients.

After bootstrapping, we find the following power law coefficients:\\
	\begin{equation*}
		\mathrm{A} = (0.404 \pm 0.8776) \cdot 10^{-7};\mathrm{ n_{LDFA} = 3.051\pm 0.201; n_{ODF}=2.337\pm 0.197}
	\end{equation*}

\begin{figure}[hbt]
  \includegraphics[width=0.66\linewidth]{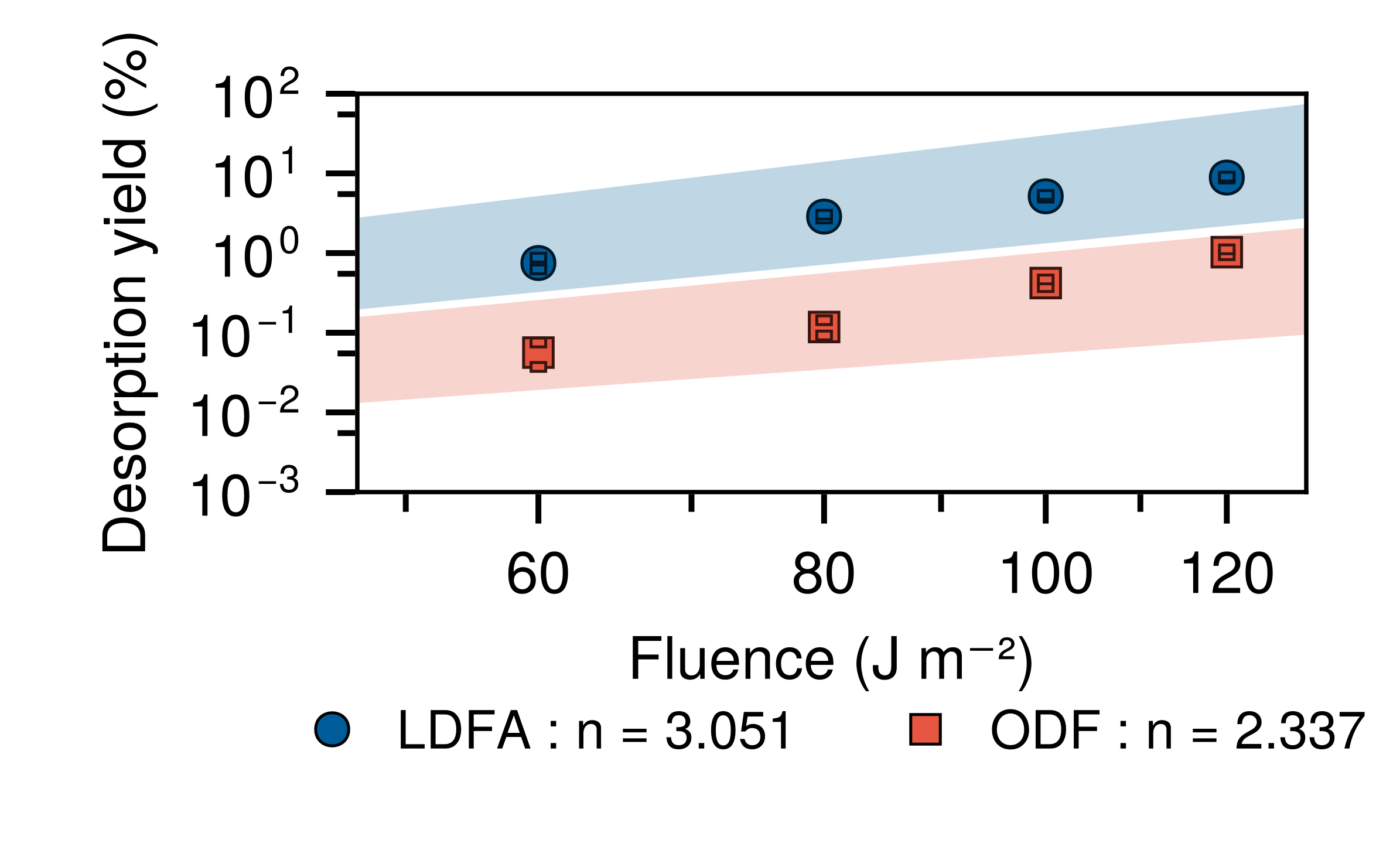}
  \caption{Desorption yields as a function of laser fluence for LDFA and ODF simulations. The indicated power law exponents $n$ are determined through bootstrapping. The bands around the desorption probability indicate 95\% confidence intervals. }
  \label{fig-yields-errorprop}
\end{figure}

The difference in power law exponents obtained for LDFA and ODF is statistically significant using the above method.
\cref{fig-yields-errorprop} shows the 95\% confidence intervals for the power law fits.

\clearpage
\section{Desorption times for laser-driven desorption simulations. }
While individual trajectories show some delay between close approach of H atoms and desorption, this is not very common across all trajectories.
We compare times to desorption between LDFA and ODF by measuring the time between close approach and desorption.

For a given trajectory, the onset of recombinative desorption is classified as the last snapshot during a trajectory where the H-H distance is larger than the distance of the H2 centre of mass from the surface.
The same condition is used to detect the transition state for desorption. The final desorbed state is classified by a H-H distance close to the equilibrium of 0.74 Å, and a centre of mass distance of at least 5 Å from the surface.

\begin{figure}[hbt]
  \includegraphics{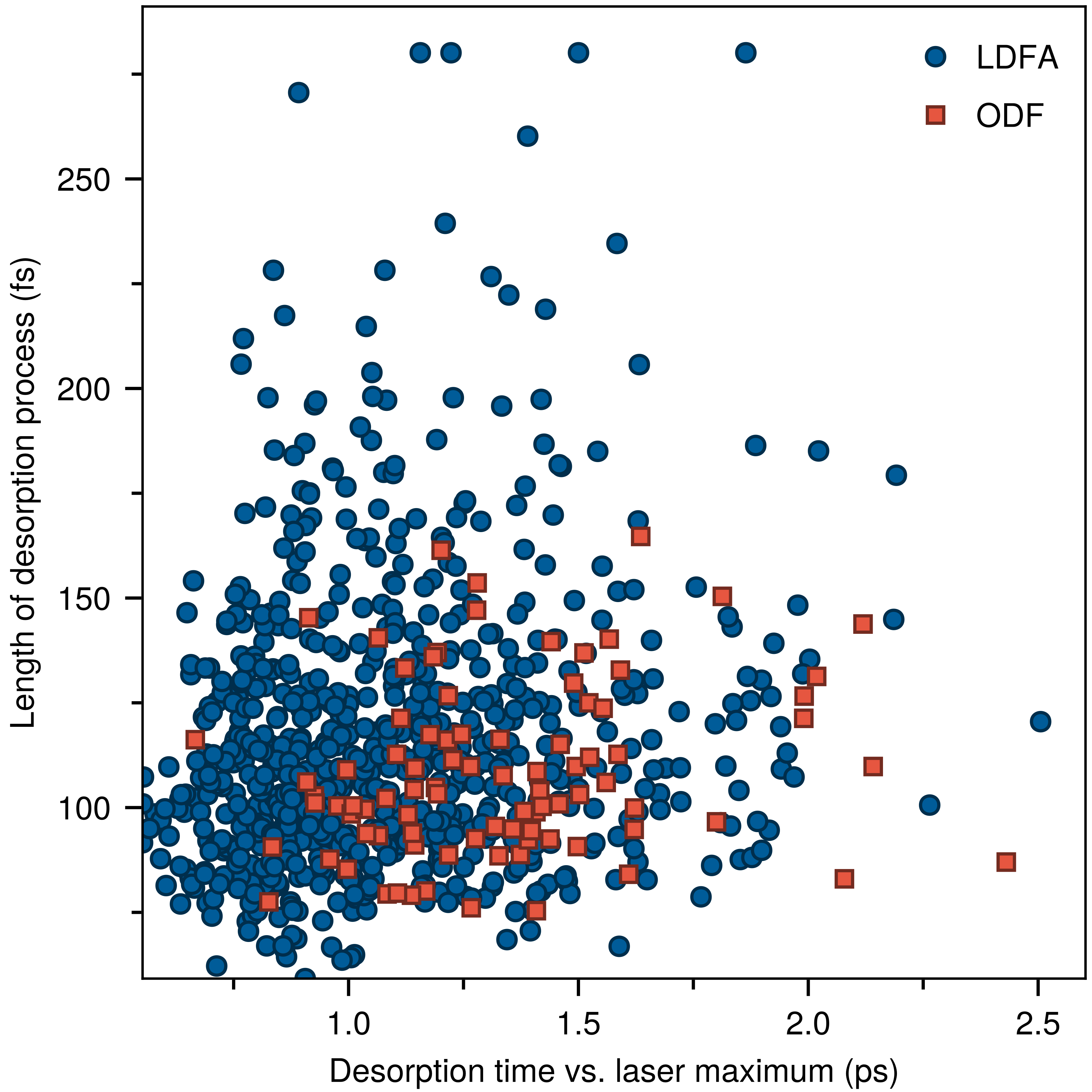}
  \caption{Correlation between the duration between onset and end of desorption with the desorption time relative to the laser pulse for LDFA and ODF simulations. }
  \label{fig-si-time-correlation}
\end{figure}

While both LDFA and ODF simulations show that the desorption process takes around 100 fs (as shown in \ref{fig-si-time-correlation}), both time distributions have some form of tail structure beside the main peak.
Due to the larger number of desorptions observed for LDFA, a larger number of outliers is visible compared to ODF.
However, we believe there would not be a significant difference if an identical number of desorption events were investigated.

\clearpage
\section{MDEF work over time for laser-driven desorption
simulations}\label{mdef-work-over-time-for-laser-driven-desorption-simulations}

Fig.~5 in the main manuscript shows the work performed by each term in the MDEF equation of motion. To quantify the importance of electronic friction, we analyse the different terms of the MDEF equation of motion, derived from eq. \ref{eq-mdef}.

The stochastic equation of motion contains \(\nabla V\), the potential
gradient informing classical dynamics. The velocity-dependent friction
term \(\mathbf \Lambda\frac{\partial}{\partial t}\mathbf{R}\) describes
the frictional force caused by nonadiabatic effects, while \(\mathcal{W}(\mathbf \Lambda, \mathrm{T_{el}}, \mathbf{R})\) describes stochastically distributed thermal random noise.

For a given simulation, we can determine the work of each force term
described above as a function of time, which allows us to estimate the
relevance of each contribution throughout the trajectory.

\[
\mathrm W(t) = \int_{\mathbf R_0}^{\mathbf R_t} \mathrm{d}\mathbf{R}~ \mathbf{F}(\mathbf R)
\]

The thermal fluctuation force is defined as follows: \[
\mathrm{F}_{\mathcal{W}}(t) = \sqrt{2\mathrm{m}\Lambda \mathrm{k_B T_{el}(t)}} * \frac{\mathrm{Normal(0,1)}}{\sqrt{\Delta \mathrm{t}}}
\]

The force of friction is defined as
\(\mathbf\Lambda\frac{\partial}{\partial t}\mathbf{R}\), where
\(\mathbf \Lambda\) is the friction tensor.

\subsection{Work components for different ODF simulations}
\begin{figure}[htbp]

{\centering \includegraphics[width=1\linewidth,height=\textheight,keepaspectratio]{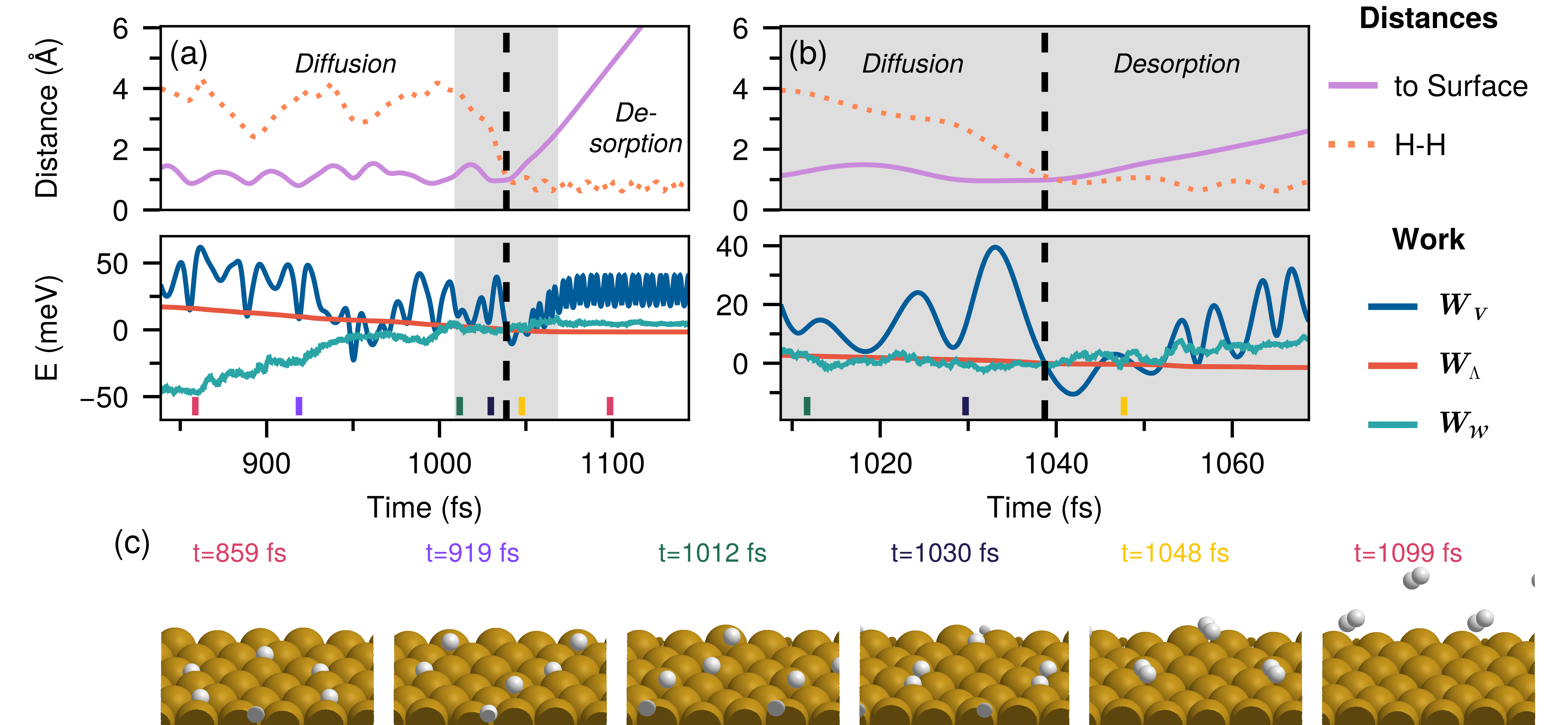}

}

\caption{Energy contributions along an ODF simulation
trajectory. This figure shows a different trajectory than shown in the main manuscript, fig.~5. \textbf{(a)}: Energy contributions by the potential energy \(\mathrm{W_V}\), friction energy
\(\mathrm{W_\Lambda}\) and thermal fluctuation energy
\(\mathrm{W_{\mathcal{W}}}\).
\textbf{(b)}: Enlarged plot of the shaded region in (a) showing energy
contributions immediately prior to laser-induced recombinative
desorption. The energy contributions are set to 0 at the onset of
desorption for clarity. The distances between hydrogen atoms and between molecule and surface are shown as an indicator of the different stages of the trajectory. MD snapshots throughout the trajectory are shown below with the times of snapshots indicated in panels a and b as vertical bars.}
\label{fig-mdef-work-example}
\end{figure}%

\begin{figure}[htbp]

{\centering \includegraphics[width=1\linewidth,height=\textheight,keepaspectratio]{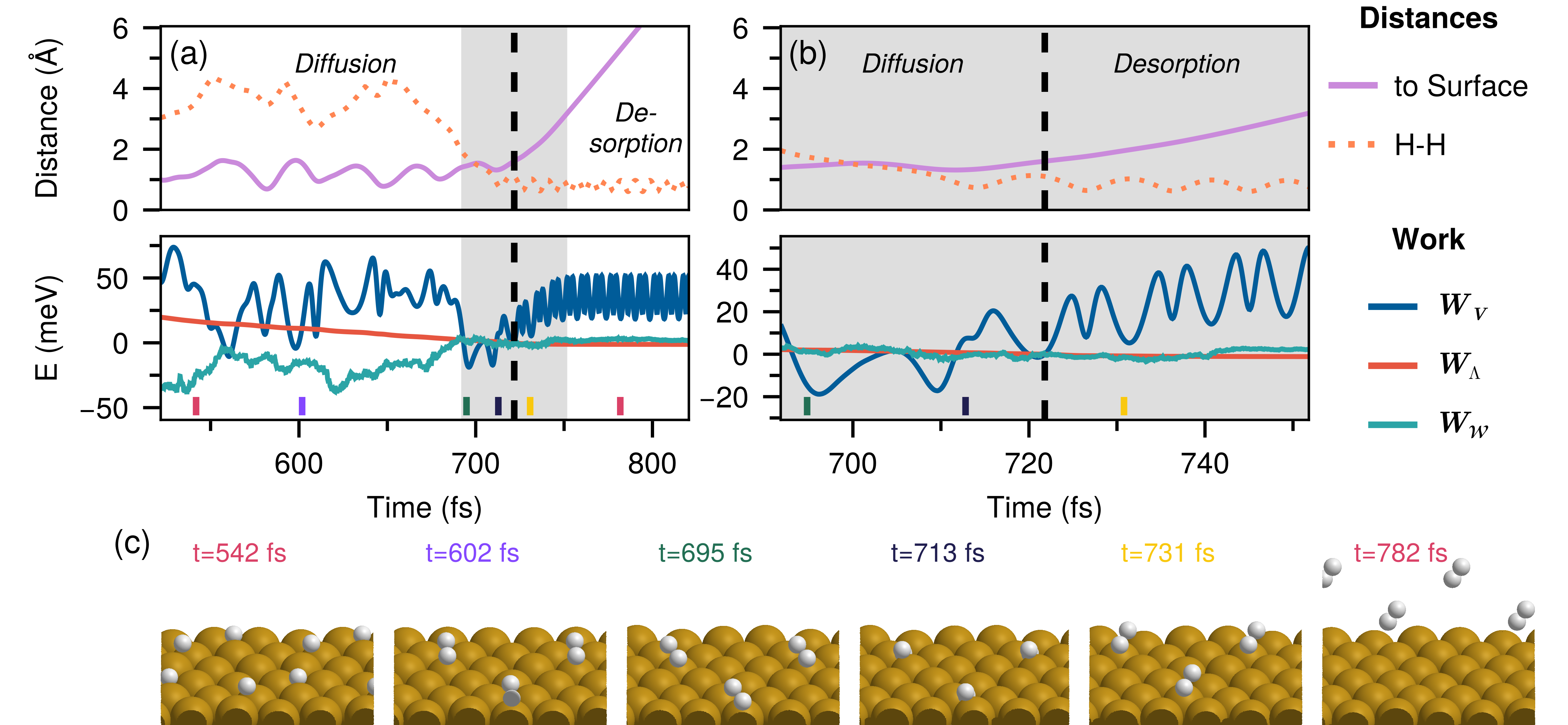}

}

\caption{Similar to fig. \ref{fig-mdef-work-example}, but for a third ODF trajectory.}

\end{figure}%

\clearpage
\subsection{Work components for different LDFA simulations}\label{ldfa-simulations}

\begin{figure}[htbp]

{\centering \includegraphics[width=1\linewidth,height=\textheight,keepaspectratio]{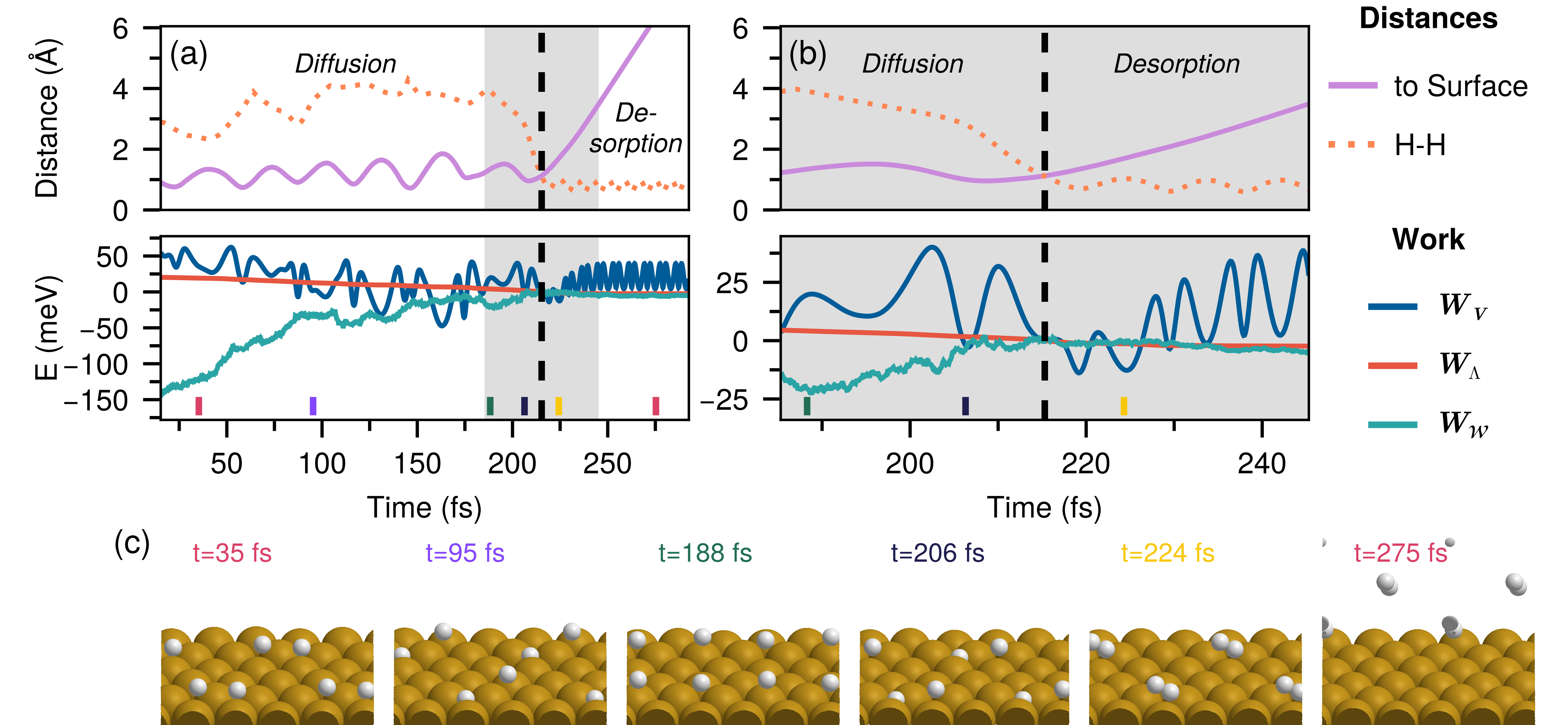}

}

\caption{Similar to \cref{fig-mdef-work-example}, but for a 2nd LDFA trajectory.}

\end{figure}%

\begin{figure}[htbp]

{\centering \includegraphics[width=1\linewidth,height=\textheight,keepaspectratio]{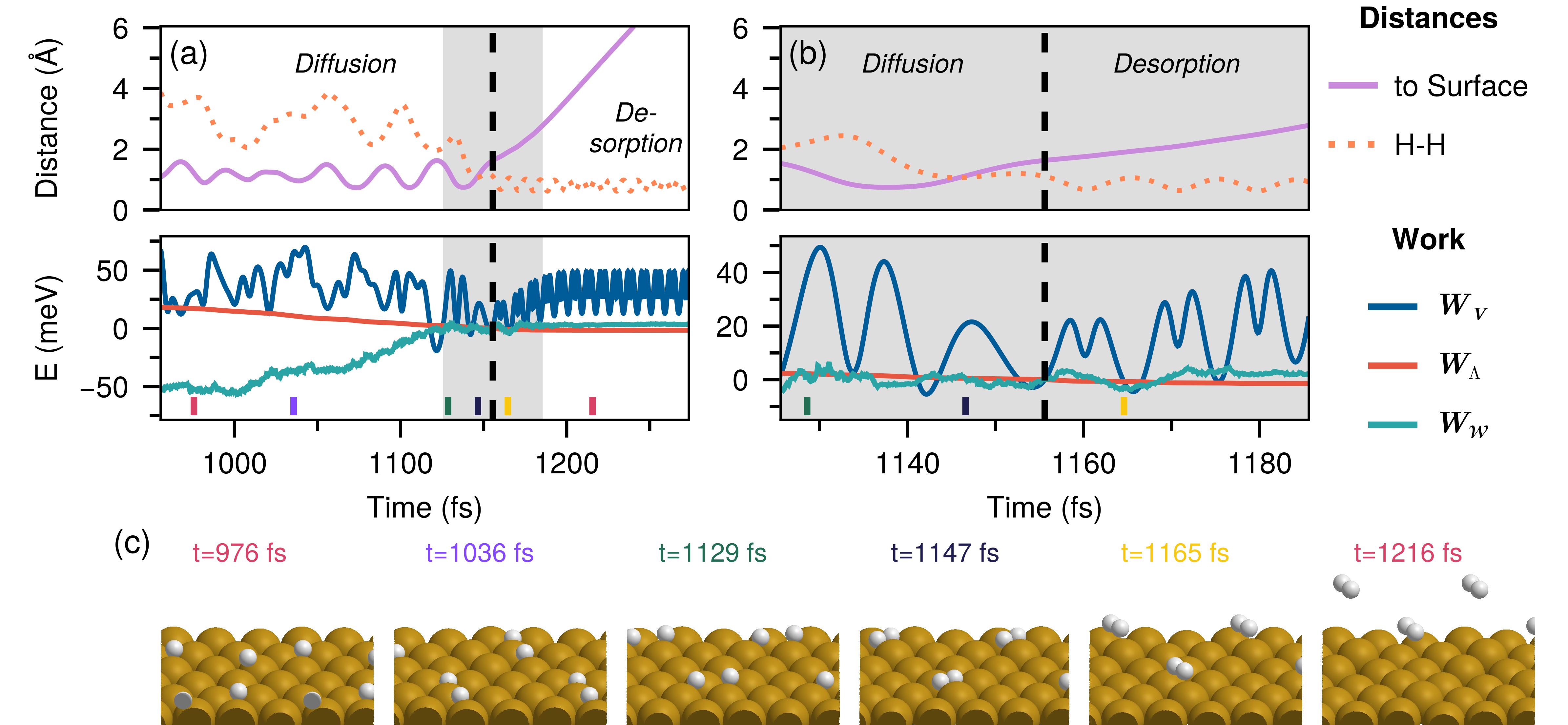}

}

\caption{Similar to \cref{fig-mdef-work-example}, but for a 3rd LDFA trajectory.}

\end{figure}%

\begin{figure}[htbp]

{\centering \includegraphics[width=1\linewidth,height=\textheight,keepaspectratio]{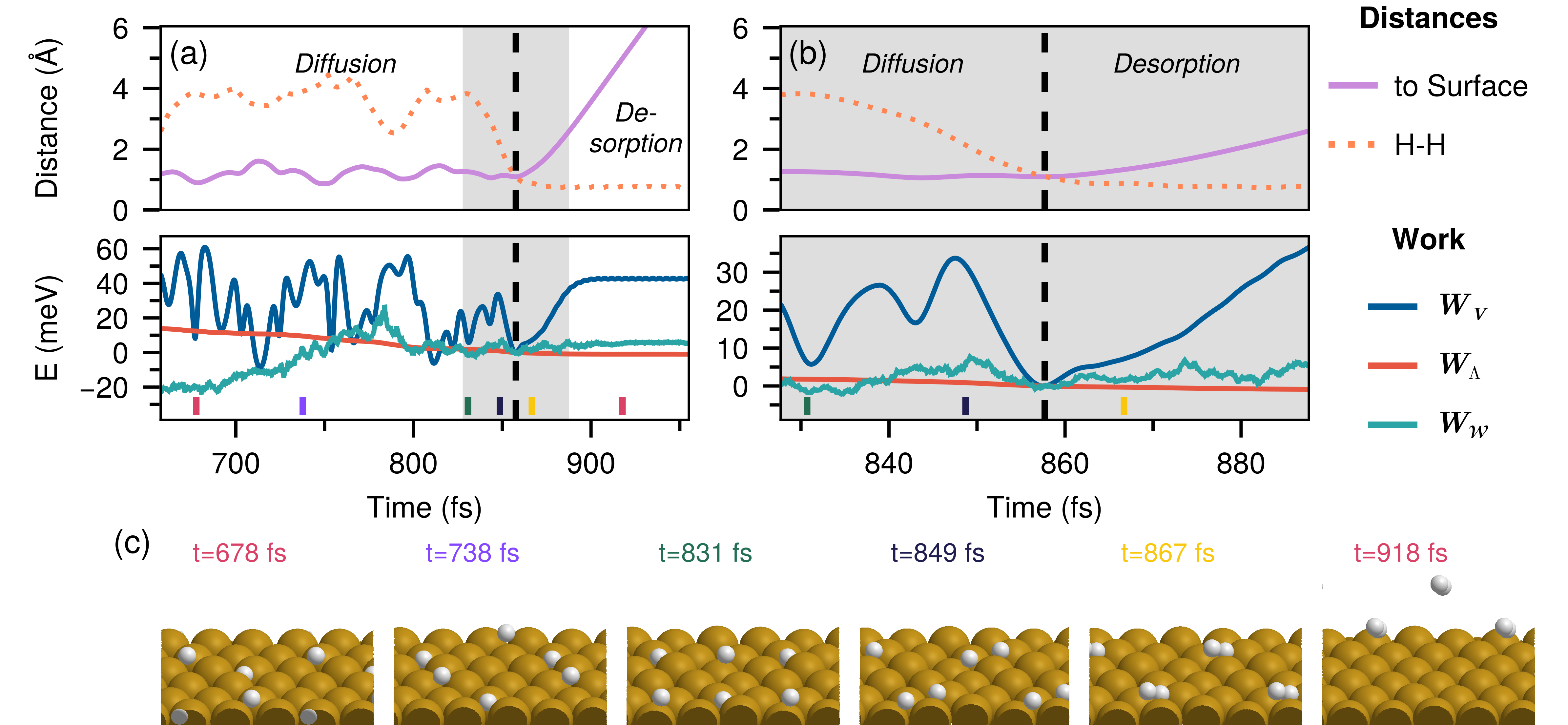}

}

\caption{Similar to \cref{fig-mdef-work-example}, but for a 4th LDFA trajectory.}

\end{figure}%

\begin{figure}[htbp]

{\centering \includegraphics[width=1\linewidth,height=\textheight,keepaspectratio]{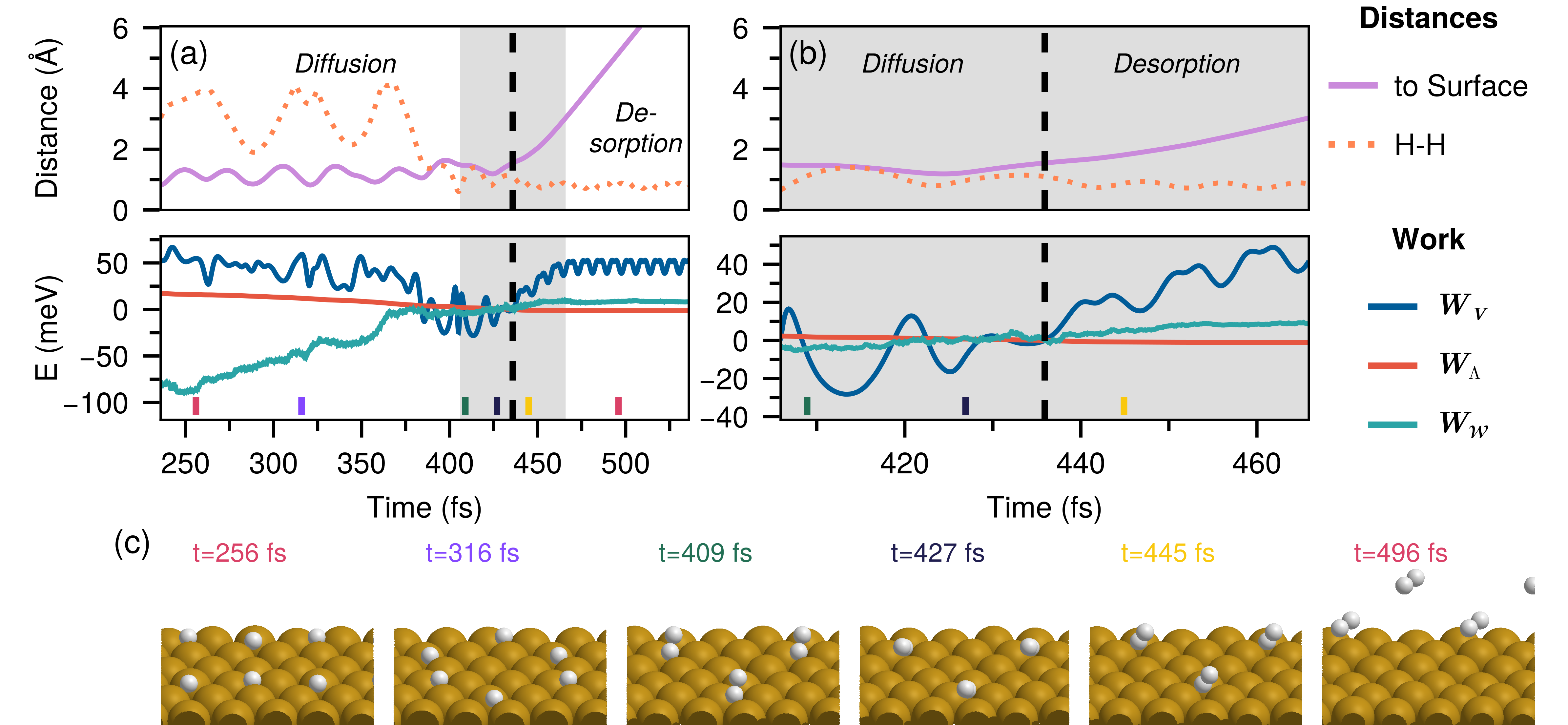}

}

\caption{Similar to \cref{fig-mdef-work-example}, but for a 5th LDFA trajectory.}

\end{figure}%

\clearpage

\bibliography{references.bib}
\typeout{get arXiv to do 4 passes: Label(s) may have changed. Rerun}